\documentclass{aa}
\usepackage{txfonts}
\usepackage{graphicx}
\usepackage{subcaption}
\usepackage[colorlinks=true,allcolors=blue]{hyperref}

\begin{document} 

\titlerunning{}
\authorrunning{Mountrichas et al.}
\titlerunning{The link among X-ray spectral properties, AGN structure and the host galaxy}

\title{The link among X-ray spectral properties, AGN structure and the host galaxy}

\author{G. Mountrichas\inst{1}, A. Viitanen\inst{2,1,3}, F. J. Carrera\inst{1},  H. Stiele\inst{4,5}, A. Ruiz\inst{6}, I. Georgantopoulos\inst{6}, S. Mateos\inst{1}, A. Corral\inst{1}}
          
    \institute {Instituto de Fisica de Cantabria (CSIC-Universidad de Cantabria), Avenida de los Castros, 39005 Santander, Spain
              \email{gmountrichas@gmail.com}
           \and
             INAF–Osservatorio Astronomico di Roma, via Frascati 33, 00040 Monteporzio Catone, Italy
             \and
             Department of Physics, University of Helsinki, PO Box 64, FI-00014 Helsinki, Finland
         \and 
            Jülich Supercomputing Centre, Forschungszentrum Jülich, Wilhelm-Johnen-Straße, 52428 Jülich, Germany
              \and
                 Institute of Astronomy, National Tsing Hua University, No. 101 Sect. 2 Kuang-Fu Road, 30013, Hsinchu, Taiwa
                 \and
                National Observatory of Athens, Institute for Astronomy, Astrophysics, Space Applications and Remote Sensing, Ioannou Metaxa
and Vasileos Pavlou GR-15236, Athens, Greece
        }

\abstract{In this work, we compare the supermassive black hole (SMBH) and host galaxy properties of X-ray obscured and unobscured AGN. For that purpose, we use $\sim 35\,000$ X-ray detected AGN in the 4XMM-DR11 catalogue for which there are available measurements for their X-ray spectral parameters, such as the hydrogen column density, N$_H$, and photon index, $\Gamma$, from the XMM2Athena Horizon 2020 European project. We construct the spectral energy distributions (SEDs) of the sources and we calculate the host galaxy properties via SED fitting analysis, utilising the CIGALE code. We apply strict photometric requirements and quality selection criteria to include only sources with robust X-ray and SED fitting measurements. Our sample consists of 1\,443 AGN. In the first part of our analysis, we use different N$_H$ thresholds (10$^{23}$\,cm$^{-2}$ or 10$^{22}$\,cm$^{-2}$), taking also into account the uncertainties associated with the N$_H$ measurements, to classify these sources into obscured and unobscured (or mildly obscured). We find that obscured AGN tend to live in more massive systems (by $\sim 0.1$\,dex) that have lower SFR (by $\sim 0.25$\,dex) compared to their unobscured counterparts. However, only the difference in stellar mass, M$_*$, appears statistically significant ($>2\sigma$). The results do not depend on the N$_H$ threshold used to classify AGN. The differences in M$_*$ and SFR are not statistically significant for luminous AGN ($\rm log\,(L_{X,2-10\,KeV}/erg\,s^{-1})> 44$). Our findings also show that unobscured AGN have, on average, higher specific black hole accretion rates, $\lambda _{sBHAR}$, compared to their obscured counterparts, a parameter which is often used as a proxy of the Eddington ratio. In the second part of our analysis, we cross-match the 1\,443 X-ray AGN with the SDSS DR16 quasar catalogue to obtain information on the SMBH properties of our sources. This results in 271 type 1 AGN, at $\rm z<1.9$. Our findings show that type 1 AGN with increased N$_H$ ($>10^{22}$\,cm$^{-2}$) tend to have higher black hole masses, M$_{BH}$, compared to AGN with lower N$_H$ values, at similar M$_*$. The M$_{BH}$/M$_*$ ratio remains consistent for N$_H$ values below 10$^{22}$\,cm$^{-2}$, but it exhibits signs of an increase at higher N$_H$ values. Finally, we detect a correlation between $\Gamma$ and Eddington ratio, but only for type 1 sources with N$_H<10^{22}$\,cm$^{-2}$.}

   
\maketitle

\section{Introduction}

Active galactic nuclei (AGN) are powered by accretion onto the supermassive black hole (SMBH) that is located in the centre of most, if not all, galaxies. It is well known that AGN and their host galaxies co-evolve. This co-eval growth is manifested in a number of ways. For instance, strong correlations have been found between the mass of the SMBH, M$_{BH}$, and various galaxy properties, such as the bulge or stellar mass, M$_*$, and the stellar velocity dispersion of galaxies \citep[e.g.,][]{Magorrian1998, Ferrarese2000, Haring2004}. Moreover, both the AGN activity and the star formation are fed by the same material (cold gas) and peak at similar cosmic times \citep[$\rm z\sim 2$; e.g.,][]{Boyle2000, Sobral2013}. Thus, for the purpose of understanding galaxy evolution, it is important to shed light on the structure of AGN, to elucidate the physical mechanism(s) that feed the central SMBH and to study the AGN feedback.

One important aspect of this pursuit it to decipher the physical difference between obscured and unobscured AGN. According to the unification model \citep[e.g.,][]{Urry1995, Nenkova2002, Netzer2015}, AGN are surrounded by a dusty gas torus structure that absorbs radiation emitted from the SMBH and the accretion disc around it. This absorbed radiation is then re-emitted at longer (infrared) wavelengths. In the context of this model, an AGN is classified as obscured or unobscured, depending on the inclination of the line of sight with respect to the symmetry axis of the accretion disk and torus. When the AGN is observed edge-on the source is characterised as obscured, while it is classified as unobscured when the AGN is viewed face-on. Recent investigations have proposed a more intricate structure for AGN to account for the diverse classifications observed at different wavelengths, such as X-ray versus optical classifications \citep[e.g.,][]{Ogawa2021, Esparza_Arredondo_2021}. Advanced techniques like infrared interferometry have enabled detailed examinations of the inner regions of AGN, unveiling complexities within the torus and challenging the conventional notion of a simple, smooth toroidal model \citep[e.g.,][]{Tristram2007}. Studies focusing on short timescale variability in X-ray column density have identified fluctuations, indicating dynamic changes in the distribution of obscuring material \citep[e.g.,][]{Marinucci2016}. While the presence of clumps in the torus contributes to the diversity of observed AGN types and offers valuable insights into the accretion processes around SMBHs, according to the unification model, the inclination angle remains the pivotal factor for distinguishing between obscured and unobscured AGN.

An alternative interpretation of the AGN obscuration comes from the class of the evolutionary models. In this case, the different AGN types are attributed to SMBH and their host galaxies being observed at different phases. The main idea of these models is that obscured AGN are observed at an early phase, when the energetic output from the accretion disk around the SMBH is not strong enough and incapable of expelling the gas that surrounds it. As material is accreted onto the SMBH, its energetic output becomes more powerful and eventually pushes away the obscuring material \citep[e.g.,][]{Ciotti1997, Hopkins2006, Somerville2008}. 

Studying the two AGN populations can shed light on many different aspects of the AGN-galaxy interplay. An important step towards this direction is to first understand what is the nature of obscured and unobscured AGN. A popular approach to answer this question is to compare the host galaxy properties of the two AGN types. If the different AGN populations live in similar environments, this would provide support to the unification model, whereas if they reside in galaxies of different properties, it would suggest that they are observed at different evolutionary phases.

Previous studies that compared the host galaxy properties of obscured and unobscured AGN found conflicting results, depending on the wavelengths used to identify AGN (optical, infrared, X-rays) and the obscuration criteria applied to classify sources. For X-ray selected AGN, when the source classification is based on optical criteria, for instance using optical spectra to identify broad and narrow emission lines, most studies have found that obscured AGN (type 2) tend to live in more massive galaxies compared to their unobscured (type 1) counterparts. However, both AGN types live in systems with similar levels of star formation \citep[e.g.,][]{Zou2019, Mountrichas2021b}. When the classification is based on X-ray criteria, though, for instance on the value of the hydrogen column density, N$_H$, some studies have not observed significant differences in the host galaxy properties of obscured and unobscured AGN \citep[e.g.,][]{Masoura2021, Mountrichas2021c}, while other studies found an increase of M$_*$ with N$_H$ \citep{Lanzuisi2017}. Differences regarding both the M$_*$ and the stellar populations of the two AGN classes have also been reported \citep{Georgantopoulos2023}.

Another important aspect that can shed light on the structure of AGN is the study of the possible correlation between the X-ray spectral index, $\Gamma$, and the Eddington ratio, n$_{Edd}$. n$_{Edd}$ is defined as the ratio of the AGN bolometric luminosity, L$_{bol}$, and the Eddington luminosity, L$_{Edd}$ (L$_{Edd}=1.26 \times 10^{38}\rm M_{BH}/M_\odot\,erg\,s^{-1}$, where M$_{BH}$ is the mass of the SMBH). The relation between $\Gamma$ and n$_{Edd}$ has been interpreted as the link between the accretion efficiency in the accretion disk and the physical status of the corona. Due to the larger amount of optical and UV photons produced by the accretion disc, the X-ray emitting plasma is more efficiently cooled at higher n$_{Edd}$ \citep[e.g.,][]{Vasudevan2007, Davis2011}. Alternatively, a correlation between $\Gamma$ and n$_{Edd}$ could be explained by the significant dependence of the cut-off energy on the n$_{Edd}$ \citep[for more details see][]{Ricci2018}. Apart from the physical interpretation of a possible correlation between $\Gamma$ and n$_{Edd}$, it is still not clear whether such a correlation is robust or universal, since relevant studies have found conflicting results \citep[e.g.,][]{Shemmer2008, Risaliti2009, Sobolewska2009, Brightman2013, Trakhtenbrot2017, Kamraj2022}.

In this work, we utilize X-ray AGN from the 4XMM-DR11 catalogue, for which there are available measurements for their X-ray spectral parameters, within the framework of the XMM2Athena{\footnote{http://xmm-ssc.irap.omp.eu/xmm2athena/}} project. The 4XMM-DR11 dataset proves advantageous for our investigations owing to its extensive collection of X-ray sources detected by the XMM-Newton observatory. The utilization of XMM-Newton is particularly beneficial due to its high sensitivity and spatial resolution, allowing for the identification of faint and distant sources. Furthermore, the dataset demonstrates homogeneity, as the processed data within it adhere to a consistent processing methodology. Using the calculations for the spectral parameters, we classify sources into X-ray obscured and unobscured and compare their host galaxy properties. We then cross-match the X-ray dataset with the \cite{Wu2022} catalogue that provides calculations for SMBH properties, such as M$_{BH}$ and L$_{bol}$, for SDSS quasars. This enables us to compare the SMBH properties of the two AGN populations. The paper is structured as follows. In Sect. \ref{sec_data} the two aforementioned catalogues are described. Sect. \ref{sec_analysis}, presents the spectral energy distribution (SED) fitting analysis followed to measure the host galaxy properties. The selection criteria and the final samples are described in Sect. \ref{sec_final_samples}. Our results and conclusions are presented in Sect. \ref{sec_results}. In Sect. \ref{sec_conclusions}, we summarise the main findings of this work.

\section{Data}
\label{sec_data}

In this work, we used X-ray AGN included in the 4XMM-DR11 catalogue \citep{Webb2020}. 4XMM-DR11 is the fourth generation catalogue of serendipitous X-ray sources from the European Space Agency's (ESA) XMM-Newton observatory. It has been created by the XMM-Newton Survey Science Centre (SSC) on behalf of ESA. The catalogue contains 319\,565 detections with spectra in 11\,907 observations. There are 100\,237 (31.4\%) detections that are made in both detectors, 135\,342 (45.5\%) detections that come solely from the pn detector and 73\,986 (23.2\%) detections that are only from the MOS cameras. 

There are 210\,444 sources that have available spectrum from one or more detections. The process followed for the classification and fitting of these sources has been done within the framework of the XMM2Athena project and is described in detail in Viitanen et al. (in prep.). In brief, the \cite{Tranin2022} catalogue was used to classify the sources. In \cite{Tranin2022}, they used a reference sample constructed by cross-matching the {\it{Swift}}-XRT \citep[2SXPS;][]{Evans2020} sample and the data release 10 of the XMM-Newton serendipitous dataset (4XMM-DR10) with catalogues of identified AGN, stars and other X-ray source types (for more details, see their Table 2 and Sect. 2.1.2). The 210\,444 sources were cross matched with the \cite{Tranin2022} catalogue, using a matching radius of 1\arcsec. The final number of sources with available classification is 92\,238. Out of them, 76\,610 are AGN, 14\,308 are stars, 1\,091 are X-ray binaries (XRBs) and 229 are cataclysmic variables (CVs).

For the X-ray sources classified as AGN, we calculated photometric redshifts for those with a reliable optical
counterpart in SDSS or PanSTARRS, following the methodology presented in \cite{Ruiz2018}. In the case of SDSS
sources with spectroscopic observations, the corresponding redshift was used instead. There are 35\,538 AGN with available redshift (8\,467 of them have spectroscopic redshift). A detailed presentation of this photometric redshift catalogue is presented in Ruiz et al. (in prep.).


The X-ray spectral analysis of these AGN was conducted within the framework of the XMM2Athena project \citep{Webb2023}. Details of this analysis can be found in Viitanen et al. (in prep.). In summary, a Bayesian spectral fit is performed with the \texttt{BXA} tool \citep{Buchner2014}, that facilitates the connection between \texttt{XSPEC}, that is used for the analysis of the X-ray spectral data \citep{Arnaud1996} and the nested sampling package \texttt{UltraNest} \citep{Buchner2021}. An uninformative prior was assigned to each parameter within the model, and the exploration of the entire parameter space, with equal-weighted sampling points, was carried out using the MLFriends algorithm \citep{Buchner2019} implemented within \texttt{UltraNest}. \texttt{UltraNest} is a Bayesian inference library designed for high-dimensional parameter estimation and model selection. \texttt{UltraNest} employs a nested sampling algorithm that works by enclosing high likelihood regions within a set of nested iso-likelihood contours. The algorithm then successively samples from the prior within these contours, effectively exploring the parameter space in a way that efficiently allocates computational resources to regions of higher likelihood. This makes nested sampling particularly effective for problems with multiple modes, as it naturally discovers and characterizes these modes. As part of the fitting process, the empirical background model in the {\sc bxa.sherpa.background} module was utilised. This model has two parts. One part models the detector background and contains the contribution of lines and is not folded through the instrumental response. The second part accounts for emission of the cosmic X-ray background and X-ray emission of the local hot bubble and Galactic halo. The background model, which is made up of a mixture of power laws, Gaussian lines and thermal emission components, is fitted in a multi-step process to the background spectrum with an increasing complexity of the model in each step, and with the parameters of the previous steps constrained within a narrow range of the best-fit value obtained in the previous step. A $\chi ^2$ test is used to estimate the goodness of fit of the background spectra. $\rm p-values$ were then obtained based on the $\chi ^2$ values and the number of effective parameters.

\begin{figure}
\centering
  \includegraphics[width=0.95\linewidth, height=7cm]{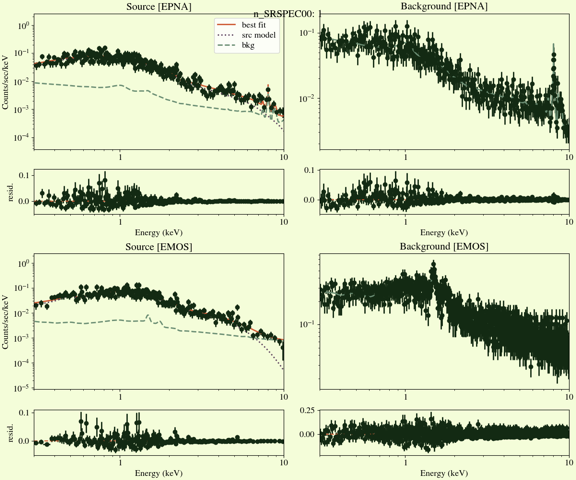}
  \caption{Example of a source spectrum and model fit (left panels), including the background modeling (right panels) for the PN (top panels) and MOS (bottom panels).}
  \label{fig_example}
\end{figure}

For the fitting of the source X-ray spectrum, a powerlaw with two absorbing media was used: the local Galactic absorption with column density fixed to the total column density in that line of sight, plus in-situ absorption at the AGN redshift with free column density. For the free parameters of the model, the following priors are applied: $\rm logf_X \in -17, -7 \,(erg/s/cm^2), \Gamma \in 0, 6, log\,NH \in 20, 26\,(cm^{-2}), and\, IIN\,(normalisation) \in 0, 5$, where $\rm f_X$ and IIN are the X-ray flux and the inter-instrument normalization (in the case of PN and MOS detections), respectively. To quantify the goodness of fit, a $\rm p-value$ is provided in the catalogue. Its calculation is described in detail in Viitanen et al. (in prep.). For each free parameter the catalogue includes the median, mode values, the 5 and 95 per cent percentiles as well as the values that encompass the narrowest 90\% interval (taken as the uncertainty of the mode). The parameters used in our analysis are the mode values of N$_H$, $\Gamma$ and the fluxes (calculated in the 2-10\,keV energy band). Using these estimated parameters, the X-ray luminosities, L$_X$, are calculated after correcting for the fitted intrinsic absorption. We opted for mode values because they are the most probable values of the distribution, capturing better than the median the cases in which they are closer to the edges of the sampled interval. It is important to highlight that this selection, however, does not impact the overall results and conclusions of our study. An example of a source spectrum and model fit, including the background modeling is presented in Fig. \ref{fig_example}.

One goal of this study is to compare the SMBH properties of X-ray obscured and unobscured AGN. To add this information to our sources, we cross-matched the 35\,538 with the \cite{Wu2022} catalogue. This catalogue includes the continuum and emission line properties for 750\,414 broad-line quasars in the data release 16 of  SDSS (DR16Q), measured from optical spectroscopy. The quasars span a range of $\rm 0.1\leq z\leq 6$ and  $\rm 44\leq log\,(L_{bol}/erg\,s^{-1})\leq 48$. L$_{bol}$ were calculated using the measured continuum luminosity at rest-frame wavelengths of 5100\,\AA, 3000\,\AA\,and 1350\,\AA, depending on the redshift of the source \citep[for more details see Sect. 4.1 in][]{Wu2022}. The catalogue also includes single-epoch virial M$_{BH}$. M$_{BH}$ have been calculated adopting the fiducial recipes on H\,$\beta$ (for sources with $\rm z<0.7$) or the Mg~{\sc ii} (for sources with $\rm 0.7\leq z<1.9$) lines. At higher redshifts the C~{\sc iv} line was measured. These M$_{BH}$ and L$_{bol}$ were used to calculate the n$_{Edd}$. 

The default redshifts in DR16Q are mostly based on the redshifts derived by the SDSS pipeline \cite{Bolton2012}. \cite{Lyke2020} also visually inspected a small fraction of quasars and updated their default redshift in DR16Q with visual redshifts. In the catalogue provided by \cite{Wu2022}, they presented improved systemic redshifts, using the measured line peaks and correcting for velocity shifts of various lines with respect to the systemic velocity (for more details see their Sect. 4.2).

\section{Calculation of the host galaxy properties}
\label{sec_analysis}

\begin{table*}
\caption{Models and the values for their free parameters used by CIGALE for the SED fitting.} 
\centering
\setlength{\tabcolsep}{1.mm}
\begin{tabular}{cc}
       \hline
Parameter &  Model/values \\
        \hline
\multicolumn{2}{c}{Star formation history: delayed model and recent burst} \\
Age of the main population & 1000, 2000, 3000, 4000, 5000, 7000, 10000, 12000 Myr \\
e-folding time & 300, 1000, 3000, 7000, 10000 Myr \\ 
Age of the burst & 50 Myr \\
Burst stellar mass fraction & 0.0, 0.005, 0.01, 0.015, 0.02, 0.05, 0.10, 0.15 \\
\hline
\multicolumn{2}{c}{Simple Stellar population: Bruzual \& Charlot (2003)} \\
Initial Mass Function & Chabrier (2003)\\
Metallicity & 0.02 (Solar) \\
\hline
\multicolumn{2}{c}{Galactic dust extinction} \\
Dust attenuation law & Charlot \& Fall (2000) law   \\
V-band attenuation $A_V$ & 0.2, 0.3, 0.4, 0.5, 0.6, 0.7, 0.8, 0.9, 1, 1.5, 2, 2.5, 3, 3.5, 4 \\ 
\hline
\multicolumn{2}{c}{Galactic dust emission: Dale et al. (2014)} \\
$\alpha$ slope in $dM_{dust}\propto U^{-\alpha}dU$ & 2.0 \\
\hline
\multicolumn{2}{c}{AGN module: SKIRTOR} \\
Torus optical depth at 9.7 microns $\tau _{9.7}$ & 3.0, 7.0 \\
Torus density radial parameter p ($\rho \propto r^{-p}e^{-q|cos(\theta)|}$) & 1.0 \\
Torus density angular parameter q ($\rho \propto r^{-p}e^{-q|cos(\theta)|}$) & 1.0 \\
Angle between the equatorial plan and edge of the torus & $40^{\circ}$ \\
Ratio of the maximum to minimum radii of the torus & 20 \\
Viewing angle  & $30^{\circ}\,\,\rm{(type\,\,1)},70^{\circ}\,\,\rm{(type\,\,2)}$ \\
AGN fraction & 0.0, 0.1, 0.2, 0.3, 0.4, 0.5, 0.6, 0.7, 0.8, 0.9, 0.99 \\
Extinction law of polar dust & SMC \\
$E(B-V)$ of polar dust & 0.0, 0.2, 0.4 \\
Temperature of polar dust (K) & 100 \\
Emissivity of polar dust & 1.6 \\
\hline
\multicolumn{2}{c}{X-ray module} \\
AGN photon index $\Gamma$ & 1.8 \\
Maximum deviation from the $\alpha _{ox}-L_{2500 \AA}$ relation & 0.2 \\
LMXB photon index & 1.56 \\
HMXB photon index & 2.0 \\
\hline
\label{table_cigale}
\end{tabular}
\tablefoot{For the definition of the various parameter, see section \ref{sec_analysis}.}
\end{table*}

For the calculation of the SFR and M$_*$ of the AGN host galaxies, we applied spectral energy distribution (SED) fitting, using the CIGALE algorithm \citep{Boquien2019, Yang2020, Yang2022}. CIGALE allows the inclusion of the X-ray flux in the fitting process and has the ability to account for the extinction of the UV and optical emission in the poles of AGN \citep{Yang2020, Mountrichas2021a, Mountrichas2021b, Buat2021}.

We used the same templates and parametric grid in the SED fitting process as those used in recent works \citep{Mountrichas2021c, Mountrichas2022a, Mountrichas2022b, Mountrichas2023, Mountrichas2023c}. In brief, the galaxy component was modelled using a delayed SFH model with a function form $\rm SFR\propto t \times exp(-t/\tau)$. A star formation burst was included \citep{Malek2018, Buat2019} as a constant ongoing period of star formation of 50\,Myr. Stellar emission was modelled using the single stellar population templates of \cite{Bruzual_Charlot2003} and was attenuated following the \cite{Charlot_Fall_2000} attenuation law. To model the nebular emission, CIGALE adopts the nebular templates based on \cite{VillaVelez2021}. The emission of the dust heated by stars was modelled based on \cite{Dale2014}, without any AGN contribution. The AGN emission was included using the SKIRTOR models of \cite{Stalevski2012, Stalevski2016}. CIGALE has the ability to model the X-ray emission of galaxies. In the SED fitting process, the intrinsic L$_X$ in the $2-10$\,keV band were used. The parameter space used in the SED fitting process is shown in Table \ref{table_cigale}.

\section{Selection criteria and final samples}
\label{sec_final_samples}

From the 35\,538 AGN available in our catalogue, we used those with $\rm flag=0$, provided by the catalogue produced by the XMM2Athena. A detailed description of the flag numbering is given in Viitanen et al. (in prep.). In brief, sources with  $\rm flag=0$ have background and source fits with a $\rm p-value>0.01$, which is the threshold used to classify a fit as acceptable. There are 29\,509 AGN that meet this criterion. In our analysis, we need reliable estimates of the host galaxy properties via SED fitting measurements. Therefore, we required all our X-ray AGN to have available photometry from SDSS or Pan-STARRS, 2MASS and WISE in the following bands: $(u), g, r, i, z, (y), J, H, K$, W1, W2, and W4 \citep[e.g.,][]{Mountrichas2021b, Mountrichas2021c, Buat2021, Mountrichas2022a, Mountrichas2022b}. There are 2,501 AGN that fulfil these requirements. We note that, although we did not require it, all these AGN also have W3 measurements available. For these sources we apply SED fitting, using the templates and parameter space described in Sect. \ref{sec_analysis}.

To restrict our analysis of sources to those with the most reliable host galaxy measurements, we excluded badly fitted SEDs. For that purpose, we considered only sources for which the reduced $\chi ^2$, $\chi ^2_r<5$. This value has been used in previous studies \citep[e.g.][]{Masoura2018, Buat2021, Mountrichas2022a, Mountrichas2022b, Koutoulidis2022, Pouliasis2022} and is based on visual inspection of the SEDs. This requirement is met by $78\%$ of the sources. To further exclude systems with unreliable measurements of the host galaxy properties, we apply the same method presented in previous studies \citep[e.g.][]{Mountrichas2021c, Mountrichas2021b, Buat2021, Koutoulidis2022, Mountrichas2023c}. This method is based on a comparison between the value of the best model and the likelihood-weighted mean value calculated by CIGALE. Specifically, in our analysis, we only consider sources with $\rm \frac{1}{5}\leq \frac{SFR_{best}}{SFR_{bayes}} \leq 5$ and $\rm \frac{1}{5}\leq \frac{M_{*, best}}{M_{*, bayes}} \leq 5$, where SFR$\rm _{best}$ and M$\rm _{*, best}$ are the best-fit values of SFR and M$_*$, respectively, and SFR$\rm _{bayes}$ and M$\rm _{*, bayes}$ are the Bayesian values estimated by CIGALE. The SFR and M$_*$ criteria are met by $88\%$ and $97\%$ of the sources. Previous studies have employed these ranges for the selection of reliable SFR and M$_*$ estimations, leveraging the CIGALE code \citep[e.g.,][]{Mountrichas2021b, Mountrichas2021c, Buat2021, Mountrichas2022a, Mountrichas2022b, Mountrichas2022c, Koutoulidis2022, Pouliasis2022, Mountrichas2023, Mountrichas2023b}. \cite{Mountrichas2021c}, explored variations in the range, considering $0.1-0.33$ for the lower limit and $3-10$ for the upper limit. They affirmed that the outcomes remained insensitive to the specific choice of these limits. In our study, we validate this observation by investigating the impact of varying these boundaries within the specified limits. Our findings indicate that such variations result in a dataset size change of less than 5\%, underscoring that they do not introduce significant alterations to our overall results and conclusions.

We also restrict the sample used in our analysis to sources with $\rm z<1.9$. The reason is twofold. Our dataset does not include obscured sources at higher redshifts and M$_{BH}$ measurements available in the \cite{Wu2022} catalogue are not reliable at $\rm z>1.9$ (i.e., M$_{BH}$ calculated using the C~{\sc iv} broad-line are considered less reliable). Finally, we excluded sources for which their redshift in the XMM dataset differs more than 0.1 compared to the redshift quoted in the \cite{Wu2022} catalogue  (16 AGN were excluded from this criterion). This threshold is determined by considering that during the process of SED fitting, redshift values are rounded to the nearest tenth decimal place. We note that no filtering has been applied regarding the uncertainties associated with the photometric redshifts available in our catalogue. The vast majority of sources with erroneous photometric redshifts will result in poor SED fits and, consequently, be excluded by the SED quality  criteria previously described. Furthermore, we conducted an investigation to determine whether our results were impacted when restricting the analysis solely to AGN with spectroscopic redshifts, which is discussed in the following section.

There are 1\,443 X-ray AGN that fulfil these criteria. Out of these sources 1\,231 lie at $\rm z<0.7$ and 212 at $\rm 0.7\leq z<1.9$. This sample is used in the analysis presented in sections \ref{sec_host_analysis} and \ref{sec_smbh_analysis}. From the 1\,443 AGN,  344 are also included in the \cite{Wu2022} catalogue. Out of them, 271 are at $\rm z<1.9$. These 271 AGN are used in our analysis presented in Sect. \ref{sec_mbh_analysis}.

\begin{figure}
\centering
  \includegraphics[width=0.95\linewidth, height=7cm]{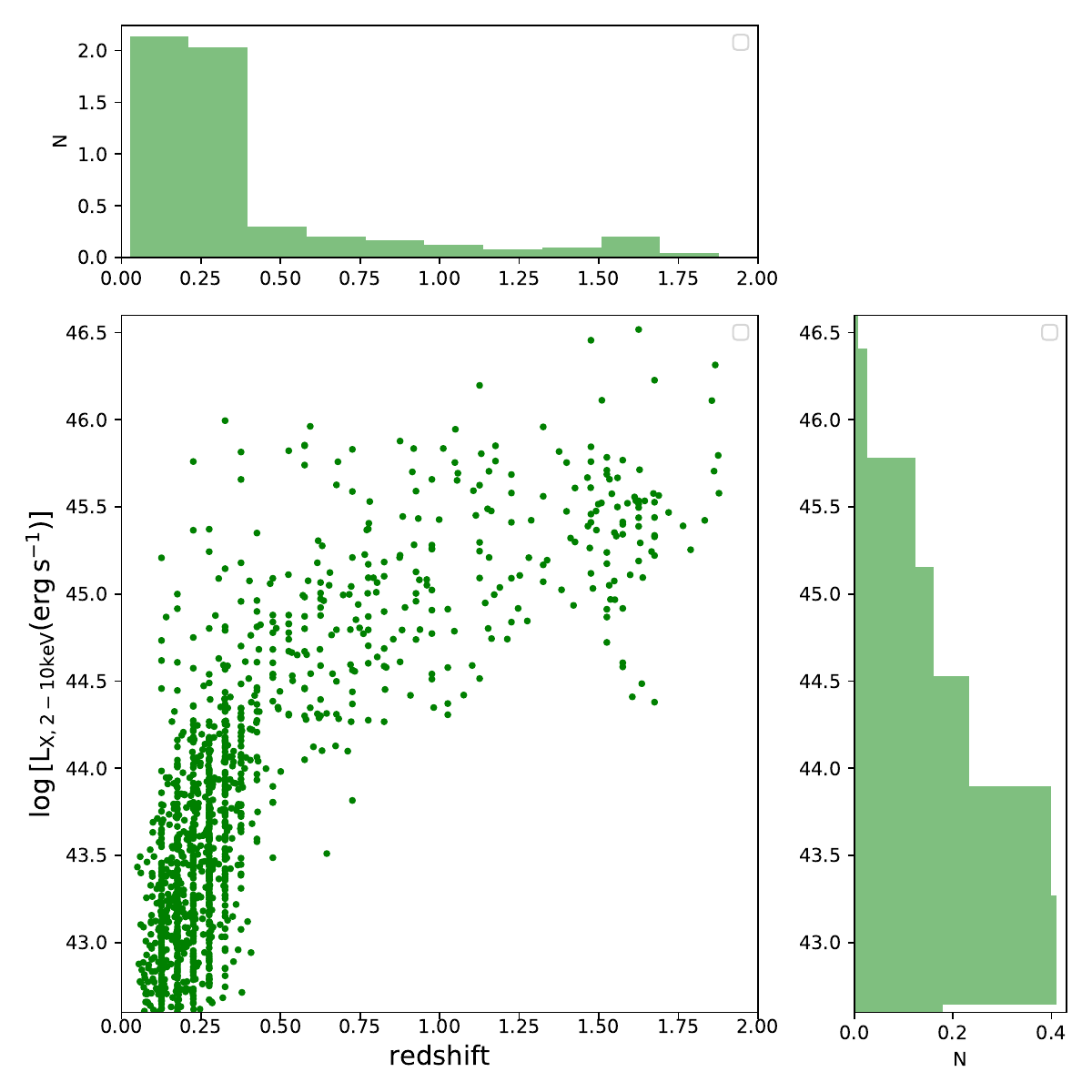}
  \caption{Distribution of the 1\,443 X-ray AGN in the L$_X-$redshift plane.}
  \label{fig_lx_redz_total}
\end{figure}

\section{Results and Discussion}
\label{sec_results}

In this section, we compare the properties of host galaxies (SFR, M$_*$) and SMBHs (n$_{Edd}$) of X-ray obscured and unobscured AGN. We also study the L$_{bol}-$M$_{BH}$ and M$_{BH}-$M$_*$ relations as a function of X-ray obscuration (N$_H$). Finally, we investigate the relation between the powerlaw slope ($\Gamma$) of the X-ray spectral model  with n$_{Edd}$.

\begin{figure}
\centering
  \includegraphics[width=0.95\linewidth, height=7cm]{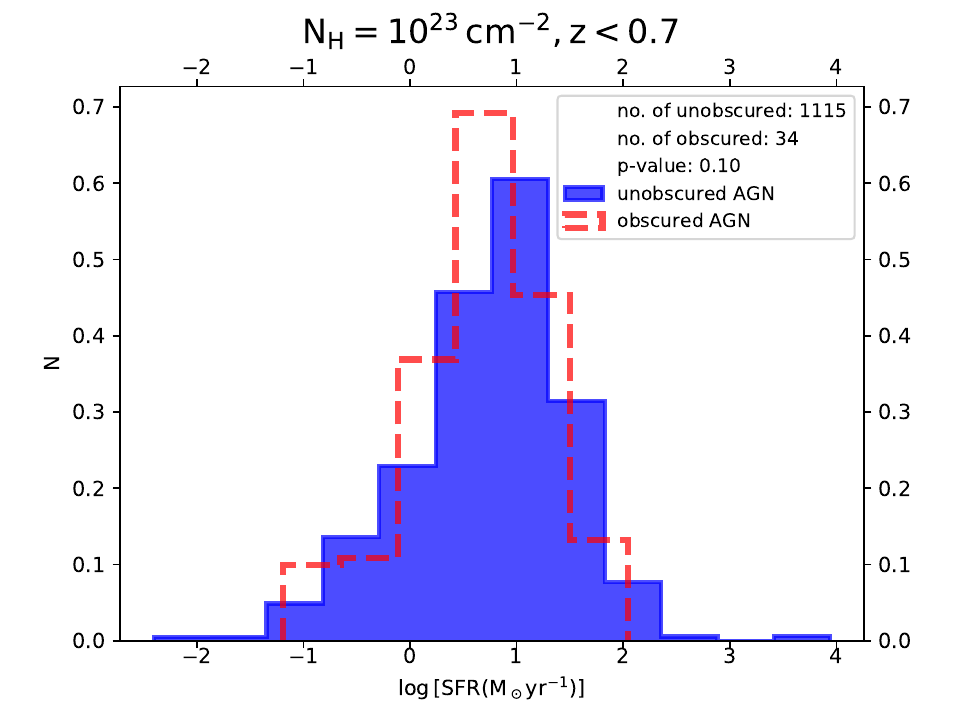}
  \includegraphics[width=0.95\linewidth, height=7cm]{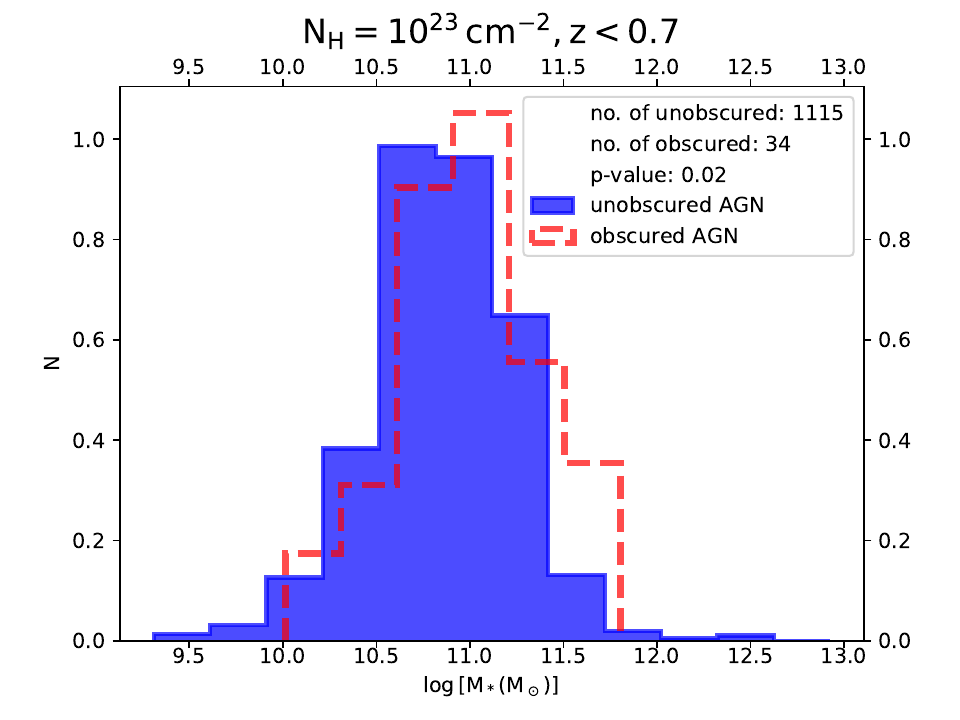}
  \caption{Host galaxy properties of X-ray obscured (red, dashed line histograms) and unobscured AGN (blue shaded histograms). Sources are classified as obscured, when the lower limit of their N$_H$ value is higher than $10^{23}$\,cm$^{-2}$. We classify AGN as unobscured if the upper limit of their N$_H$ value is below  $10^{23}$\,cm$^{-2}$. The top panel presents the SFR distributions of the two AGN populations. The bottom panel shows their M$_*$ distributions. The number of sources and the $\rm p-values$  obtained by applying a KS test, are shown in the legend of the plots. Distributions are weighted based on the redshift and L$_X$ of the sources.}
  \label{fig_sfrmstar_23}
\end{figure} 

\begin{figure}
\centering
  \includegraphics[width=0.95\linewidth, height=7cm]{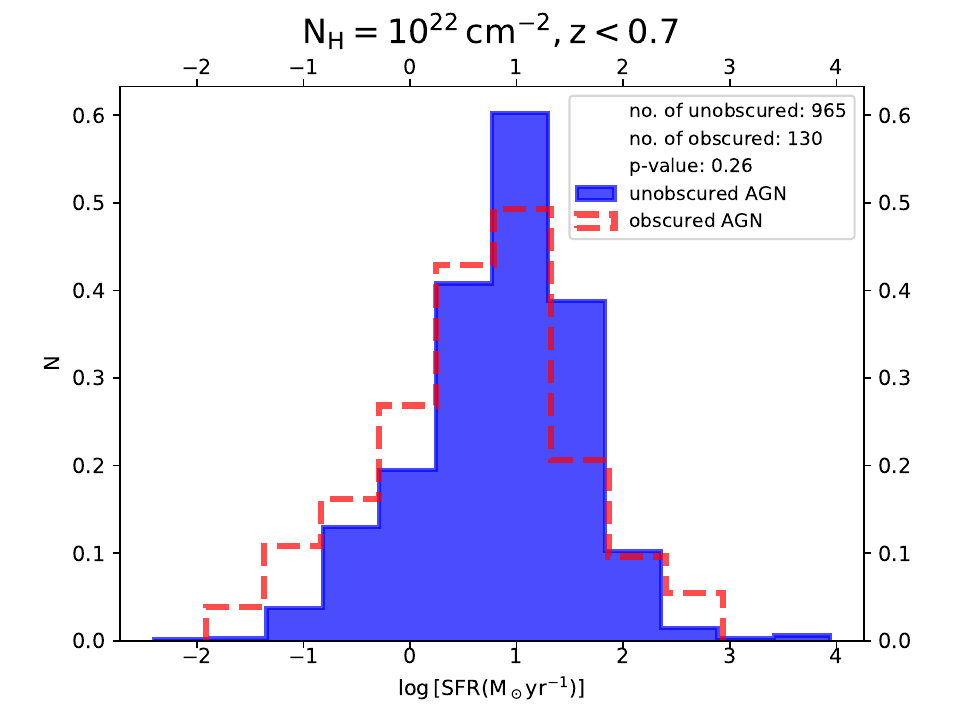}
  \includegraphics[width=0.95\linewidth, height=7cm]{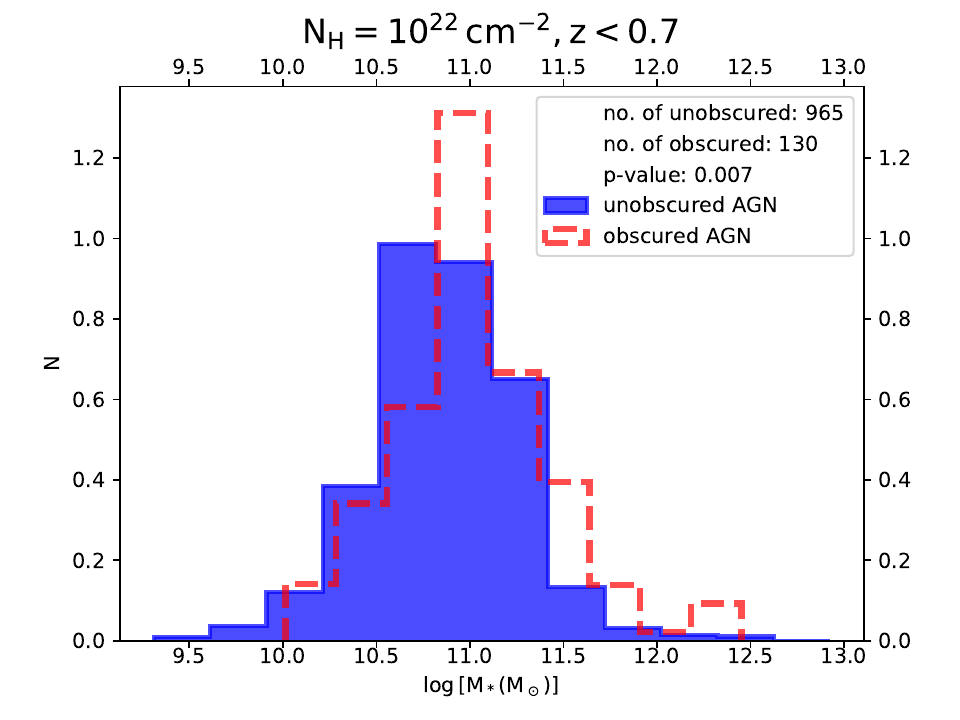}
  \caption{Similar as in Fig. \ref{fig_sfrmstar_23}, but using a limit of N$_H=10^{22}$\,cm$^{-2}$, to classify AGN (taking into account the errors of the N$_H$ measurements).} 
  \label{fig_sfrmstar_22}
\end{figure} 

\begin{figure}
\centering
  \includegraphics[width=0.95\linewidth, height=7cm]{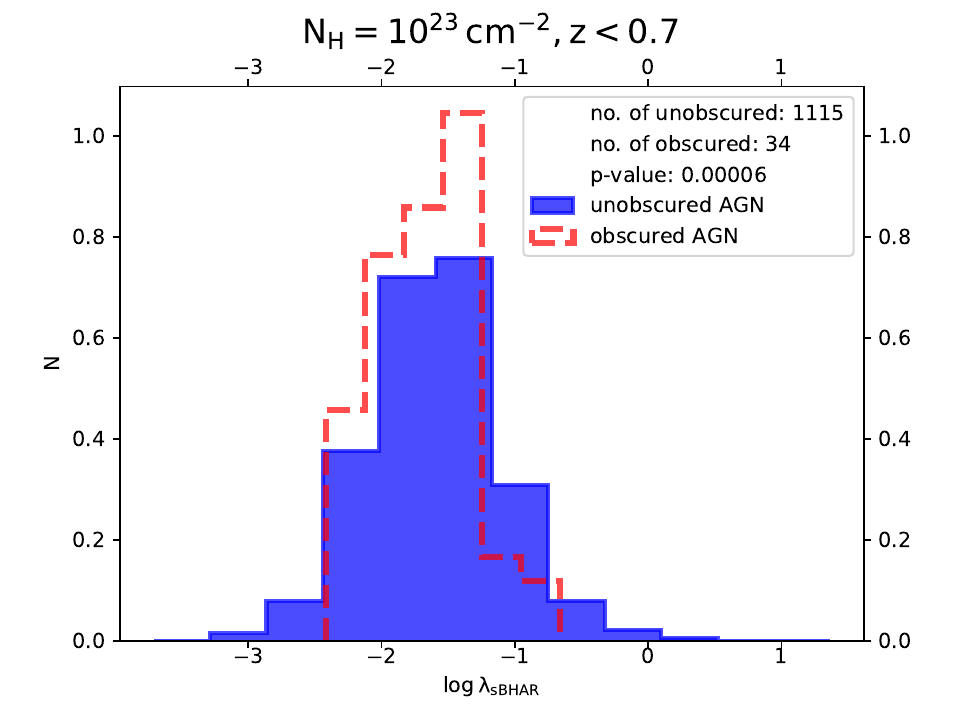}
  \includegraphics[width=0.95\linewidth, height=7cm]{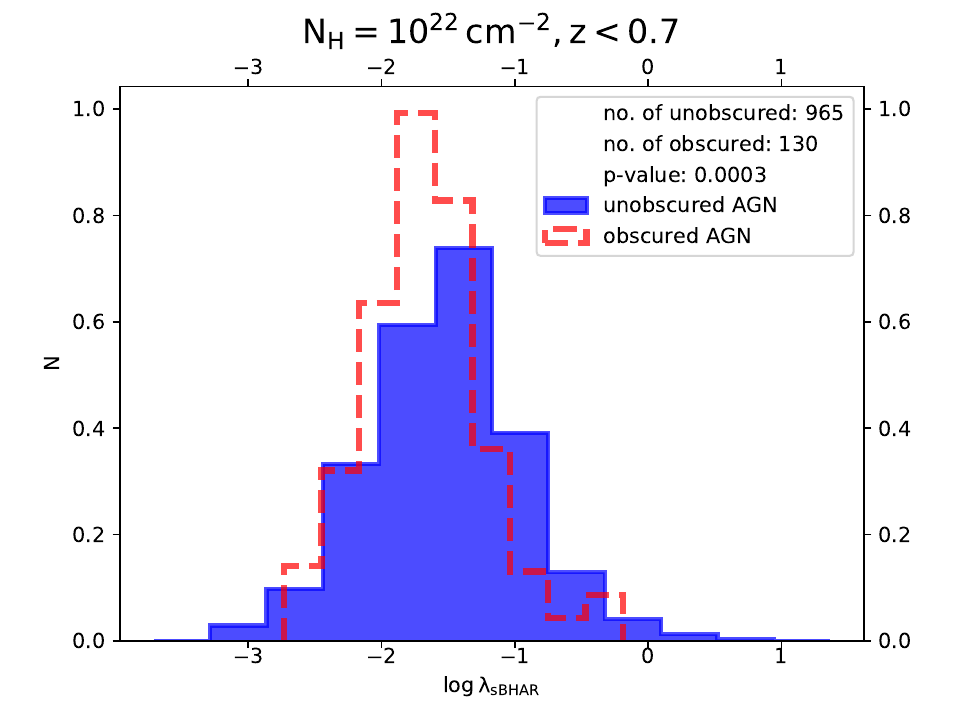}
  \caption{$\lambda _{sBHAR}$ distributions for obscured and unobscured AGN. The top panel presents the distributions using an N$_H$ threshold of 10$^{23}$\,cm$^{-2}$ for the source classification. The bottom panel presents the results using N$_H=$10$^{22}$\,cm$^{-2}$ to characterize sources. In both cases, the errors of the N$_H$ measurements have been taken into account (see text for more details). Distributions are also weighted based on the redshift and L$_X$ of the sources.} 
  \label{fig_lambda}
\end{figure}

\subsection{Host galaxy properties of X-ray obscured and unobscured AGN}
\label{sec_host_analysis}

In this part of our analysis, we use the 1\,443 AGN that meet the criteria described in Sect. \ref{sec_final_samples}. Their distribution in the L$_X-$redshift plane is presented in Fig. \ref{fig_lx_redz_total}. First, we split the sources into X-ray obscured and unobscured (or mildly obscured) AGN, using a N$_H$ limit of 10$\rm ^{23}cm^{-2}$. Specifically, we used the lower and upper limits of the 90\% confidence interval on the N$_H$ mode from the spectral fitting analysis. An AGN was classified as obscured when the lower limit of their N$_H$ value is higher than $10^{23}$\,cm$^{-2}$. Similarly, an AGN was identified as unobscured (or moderately obscured), if the upper limit of its N$_H$ value is lower than $10^{23}$\,cm$^{-2}$. Table \ref{table_numbers} presents the number of sources classified. Then, we compared the SFR and M$_*$ of the host galaxies of the two AGN populations. The selection of classification criteria serves a dual purpose. Firstly, it relies on recent research findings \citep{Georgantopoulos2023}, which highlight that distinctions in the properties of SMBHs and their host galaxies between obscured and unobscured AGN tend to diminish when the N$_H$ value falls below the threshold of N$_H<23$\,cm$^{-2}$. Secondly, by taking into account the uncertainties associated with N$_H$, the classification ensures that only firmly obscured and unobscured sources are categorised as such, enhancing the reliability of the classification process.

The results are presented in Fig. \ref{fig_sfrmstar_23}. We have restricted both datasets to $\rm z<0.7$. This is because there are no AGN classified as obscured using the criteria described above, at $\rm z>0.7$ in our dataset. We also note, that the distributions have been weighted to account for the different L$_X$ and redshift  of obscured and unobscured AGN. For that purpose, each source was assigned a weight, based on its L$_X$ and redshift, following the procedure described in e.g., \cite{Mountrichas2019, Masoura2021, Mountrichas2021c, Buat2021, Mountrichas2022c, Koutoulidis2022}. Specifically, the weight was calculated by measuring the joint L$_X$ distributions of the two populations (i.e., we add the number of obscured and unosbcured AGN in each L$_X$ bin, in bins of 0.1\,dex) and then normalise the L$_X$ distributions by the total number of sources in each bin. The same procedure was followed for the redshift distributions. The total weight that was assigned in each source was the product of the two weights. We made use of these weights in all distributions presented in the remainder of this section.

Fig. \ref{fig_sfrmstar_23}, shows the distributions of unobscured (blue shaded histograms) and obscured AGN (red, dashed line histograms). The top panel presents the SFR distributions of the two AGN classes. The (weighted) median $\rm log\,SFR$ of the unobscured and obscured AGN is 0.82 and 0.57 respectively (Table \ref{table_numbers}). Applying a Kolmogorov-Smirnov (KS) test yields a $\rm p-value$ of 0.10 (also shown in the legend of the plot). Similar $\rm p-values$  are obtained, applying a Mann-Whitney (MW, $\rm p-value=0.14$) and Kuiper ($\rm p-value=0.32$) tests. Therefore, the difference of $0.25$\,dex does not appear statistically significant (two distributions differ with a statistical significance of $\sim 2\sigma$ for a $\rm p-value$ of 0.05).

The bottom panel of Fig. \ref{fig_sfrmstar_23}, shows the M$_*$ distributions of obscured (blue, shaded histogram) and unobscured (red, dashed line) AGN. The weighted median of the $\rm log\,M_*$ is 10.86 and 10.95 for unobscured and obscured AGN, respectively. Although the difference is small, it appears to be statistically significant (Table \ref{table_numbers}). Specifically, the $\rm p-value$ derived from a KS-test is 0.02. Similar $\rm p-values$  are also obtained by applying MW- and Kuiper tests ($<0.05$). We note that the (weighted) standard deviations (SD) of the two populations are 11.22 (obscured) and 11.13 (unobscured). The (weighted) 95\% confidence intervals (CI) are $10.91-11.30$ and $10.99-11.07$, for the obscured and unobscured AGN, respectively. Therefore, based on our results, obscured AGN tend to live in more massive systems and their hosts have lower SFR compared to their unobscured counterparts. However, only the M$_*$ difference appears to be statistically significant. 

Furthermore, we explored whether our outcomes are influenced by potential redshift evolution within the $\rm z<0.7$ redshift range encompassed by our dataset. For that purpose, we split the AGN sample into two redshift bins, using the median value of the $\rm z<0.7$ dataset (i.e., at $\rm z=0.21$). The results do not appear redshift dependent, since similar trends and $\rm p-values$  were obtained. Additionally, we examined whether our findings might be sensitive to uncertainties associated with photometric redshift estimates. To address this concern, we narrowed down our analysis to include only sources with spectroscopic redshifts. We obtained similar results in this restricted subset.

Most previous studies that examined the host galaxy properties of X-ray obscured and unobscured AGN did not find (significant) differences for the SFR and M$_*$ of the hosts of the two AGN populations. Specifically, \cite{Masoura2021} and \cite{Mountrichas2021c} studied the SFR and M$_*$ of X-ray obscured and unobscured AGN at $\rm 0<z<3.5$, using data from the XMM-XXL and the Bo$\ddot{o}$tes fields, respectively, and concluded that both AGN classes live in galaxies with similar properties \citep[see also][]{Merloni2014}. We note that the aforementioned studies  did not account for the uncertainties of the N$_H$ values and applied a lower N$_H$ threshold ($\rm N_H = 10 ^{21.5}cm^{-2}$) to classify their AGN. However, \cite{Lanzuisi2017} studied AGN, at $\rm 0.1<z<4$, in the COSMOS field and found that the average N$_H$ shows a clear positive correlation with M$_*$, but not with SFR. Studies that used optical criteria (e.g., optical spectra) to classify the AGN into type 1 and 2 found that the two AGN types live in galaxies with similar SFR, but type 2 sources tend to reside in more massive systems by $\sim 0.3$\,dex, both at low redshift \citep[z<1; ][]{Mountrichas2021c} and at high redshift \citep[z<3.5;][]{Zou2019}.

More recently, \cite{Georgantopoulos2023} compared the stellar populations and M$_*$, among others, of X-ray obscured and unobscured AGN host galaxies in the COSMOS field at $\rm z\sim 1$, using $\rm N_H = 10 ^{23}cm^{-2}$, to classify their sources. Their analysis showed that the distribution of M$_*$ of obscured AGN is skewed to higher M$_*$ compared to unobscured AGN. Specifically, they found that heavily obscured AGN live in galaxies that have higher M$_*$ by 0.13\,dex compared to unobscured sources. Although the difference was small it was statistically significant ($\rm p-value=1.3\times 10^{-3}$). Their results are in excellent agreement with our findings. They also found that heavily obscured AGN tend to live in systems with older stars compared to unobscured. This is consistent with our finding for a lower SFR of heavily obscured AGN hosts compared to unobscured, although in our measurements the difference did not appear statistically significant. 


\cite{Georgantopoulos2023} suggested that a possible cause for the discrepant results among previous studies could be the different N$_H$ thresholds used. Indeed, when they lowered the N$_H$ value used to classify their AGN, the statistical significance of the differences detected in the host galaxy properties of the two AGN classes was lower (see their Fig. 13). To verify this hypothesis, we lowered the N$_H$ threshold utilised for our AGN classification to $\rm N_H = 10 ^{22}cm^{-2}$, using the lower and upper limits of N$_H$ specified at the outset of this section, and repeated the analysis. The findings are showcased in Fig. \ref{fig_sfrmstar_22} and Table \ref{table_numbers}. Similar results were emerged as those presented in Fig. \ref{fig_sfrmstar_23}. Specifically, obscured AGN tend to inhabit more massive systems (by 0.12\,dex) and have lower SFR (by 0.26\,dex) compared to unobscured AGN. However, among these differences, only the variation in M$_*$ demonstrates statistical significance. It is observed that the (weighted) standard deviation of the two populations is 11.54 for the obscured AGN and 11.29 for the unobscured AGN. The (weighted) 95\% confidence intervals are $11.10-11.40$ for the obscured AGN and $11.00-11.10$ for the unobscured AGN. Nonetheless, if we ignore the errors of the N$_H$ measurements, effectively classifying AGN based solely on their N$_H$ values without considering the limits of the measurements and we implement a threshold at $\rm N_H = 10 ^{22}cm^{-2}$, similar to what most prior studies have used for AGN classification, we uncover a marginally statistically significant different in the M$_*$ distributions of the two AGN types ($\rm p-value=0.052$) and no statistically significant difference in terms of the SFR distributions of the obscured and unobscured AGN. 

An alternative hypothesis proposed by \cite{Georgantopoulos2023} is that the variations in results across different studies could be attributed to the diverse luminosity ranges covered by the respective samples. For instance, AGN in the Bo$\ddot{o}$tes and XXM-XXL fields \citep{Masoura2021, Mountrichas2021c} probe higher luminosities compared to those probed by sources in the COSMOS field \citep{Lanzuisi2017, Georgantopoulos2023}. This is also true, when we compare the results between studies that used X-ray and optical criteria to classify the AGN. The latter have either used sources in the COSMOS field \citep[e.g.,][]{Zou2019} that, mainly, probe low to intermediate L$_X$ AGN, or they have restricted their analysis to low luminosity systems at low redshift \citep[e.g.,][]{Mountrichas2021c}. To examine this hypothesis, we restricted our X-ray dataset to luminous sources, by applying a luminosity cut at $\rm log\,[L_{X}(ergs^{-1})>44]$ (at $\rm z<0.7$) and we classified AGN using an N$_H$ value of $10^{22}$  (using the lower and upper limits of N$_H$) and repeated the analysis. There are 17 luminous and obscured and 223 luminous and unobscured AGN. We detected a difference of 0.35\,dex (based on the weighted median values) in the SFR of the two AGN populations. Obscured AGN appear to live in more massive systems compared to unobscured (by 0.06\,dex). However, none of these differences appears statistically significant ($\rm p-value= 0.52$ and 0.17, for the SFR and M$_*$, respectively).

\begin{table}
\caption{Weighted median values of SFR, M$_*$ and $\lambda _{sBHAR}$ for obscured and unobscured AGN, using different N$_H$ thresholds for the source classification. The number of sources and the p$-$values yielded by applying KS$-$tests, are also presented.}
\centering
\setlength{\tabcolsep}{1mm}
\begin{tabular}{ccccc}
 \hline
 & \multicolumn{2}{c}{N$_H=10^{23}\,\rm cm^{-2}$} &  \multicolumn{2}{c}{N$_H=10^{22}\,\rm cm^{-2}$} \\
 \hline
  & unobscured & obscured & unobscured & obscured   \\
 \hline
number of AGN & 1115 & 34 & 965 & 130  \\ 
 \hline
log\,SFR  &  0.82 & 0.57 & 0.93 & 0.67  \\
p$-$value (SFR) & \multicolumn{2}{c}{0.10} & \multicolumn{2}{c}{0.26}  \\
 \hline
log\,M$_*$ & 10.86 & 10.95 & 10.85 & 10.97  \\ 
p$-$value (M$_*$) & \multicolumn{2}{c}{0.02} & \multicolumn{2}{c}{0.007}  \\
 \hline
log\,$\lambda _{sBHAR}$ & -1.60 & -1.71 & -1.52 & -1.73  \\
p$-$vaule ($\lambda _{sBHAR}$) & \multicolumn{2}{c}{0.00006} & \multicolumn{2}{c}{0.0003} \\
  \hline
\label{table_numbers}
\end{tabular}
\end{table}

\subsection{Eddington ratio of X-ray obscured and unobscured AGN}
\label{sec_smbh_analysis}

The difference with the highest statistical significance reported by \cite{Georgantopoulos2023} was that between the n$_{Edd}$ of obscured and unobscured sources (see e.g., their Table 3). Prompted by their results, we examined this difference in our dataset. For the dataset used in this part of our analysis, we do not have available M$_{BH}$ and therefore n$_{Edd}$ measurements for our sources. Cross-matching our sample with the \cite{Wu2022} catalogue (see Sect. \ref{sec_data}) in which there are available n$_{Edd}$ measurements, would constrain our dataset to only broad line sources and exclude narrow line AGN. In the next section, we examine whether n$_{Edd}$ depends on N$_H$ (for broad line AGN). Here, to include both broad and narrow line sources in our investigation, in place of n$_{Edd}$, we examined the specific black hole accretion rate, $\lambda _{sBHAR}$ of obscured and unobscured AGN. $\lambda _{sBHAR}$ is the rate of the accretion onto the SMBH relative to the M$_*$ of the host galaxy. It is often used as a proxy of n$_{Edd}$, in particular when M$_{BH}$ measurements are not available \citep[e.g.][]{Georgakakis2017, Aird2018, Mountrichas2021c, Pouliasis2022}. For the calculation of $\lambda _{sBHAR}$ the following expression was used \citep[e.g.,][]{Aird2018}:

\begin{equation}
\lambda_{sBHAR}=\rm \frac{k_{bol}\,L_{X,2-10\,keV}}{1.26\times10^{38}\,erg\,s^{-1}\times0.002\frac{M_{*}}{M_\odot}},   
\label{eqn_lambda}
\end{equation}
where $\rm k_{bol}$ is a bolometric correction factor, that converts the X-ray luminosity to AGN bolometric luminosity. We adopted the value of $\rm k_{bol}=25$ \citep[e.g.,][]{Georgakakis2017, Aird2018, Mountrichas2021c, Mountrichas2022a}. 

Fig. \ref{fig_lambda} presents the $\lambda _{sBHAR}$ distributions of obscured and unobscured AGN, using different N$_H$ thresholds to classify sources, as indicated on top of each panel (taking into account the errors on the N$_H$ measurements). All distributions are weighted based on the L$_X$ and redshift of the sources. Table \ref{table_numbers} presents the (weighted) median $\rm log$\,$\lambda _{sBHAR}$ of the obscured and unobscured AGN, for N$_H=10^{23}$cm$^{-2}$ and N$_H=10^{22}$cm$^{-2}$. A difference of $\sim 0.1-0.2$\,dex is found in the $\lambda _{sBHAR}$ values of the two populations. Specifically, obscured sources tend to have, on average, lower $\lambda _{sBHAR}$ compared to unobscured sources. Based on the $\rm p-values$  derived using a KS-test, this different has a high statistical significance, regardless of the N$_H$ limit used. Similar $\rm p-values$ are found applying MW and Kuiper tests.


Our results are in agreement with those reported in \cite{Georgantopoulos2023}. We note that the difference we found appears smaller compared to that presented in the aforementioned study ($\sim 0.3$\,dex). As we have already pointed out, in our analysis we have used more strict criteria for the classification of AGN as obscured and unobscured compared to those used in \cite{Georgantopoulos2023}. Moreover, in \cite{Georgantopoulos2023}, for the calculation of n$_{Edd}$ they have applied the same bolometric correction on L$_X$ to infer L$_{bol}$, as in this study. However, they have used M$_{BH}$ measurements derived by stellar velocity dispersions, $\sigma$, that is they have utilised a different scaling relation compared to our work.

We conclude that obscured AGN tend to live in more massive systems compared to unobscured, at $\rm z<0.7$. The difference in M$_*$ is small ($\approx 0.1$\,dex), but appears statistically significant. Obscured sources also live in hosts that have lower SFR, by  $\approx 0.25$\,dex, than unobscured, however, this difference is not statistically significant ($< 2\,\sigma$). Moreover, neither of these differences appears to be statistically significant for luminous AGN ($\rm log\,[L_{X}(ergs^{-1})>44]$ and/or when less strict criteria are used for the source classification (i.e., errors on N$_H$ are ignored). Based on our analysis, the difference between the two AGN populations, with the highest statistical significance is that regarding the $\lambda _{sBHAR}$, that is a proxy of the n$_{Edd}$.

\begin{figure}
\centering
  \includegraphics[width=0.99\linewidth, height=7cm]{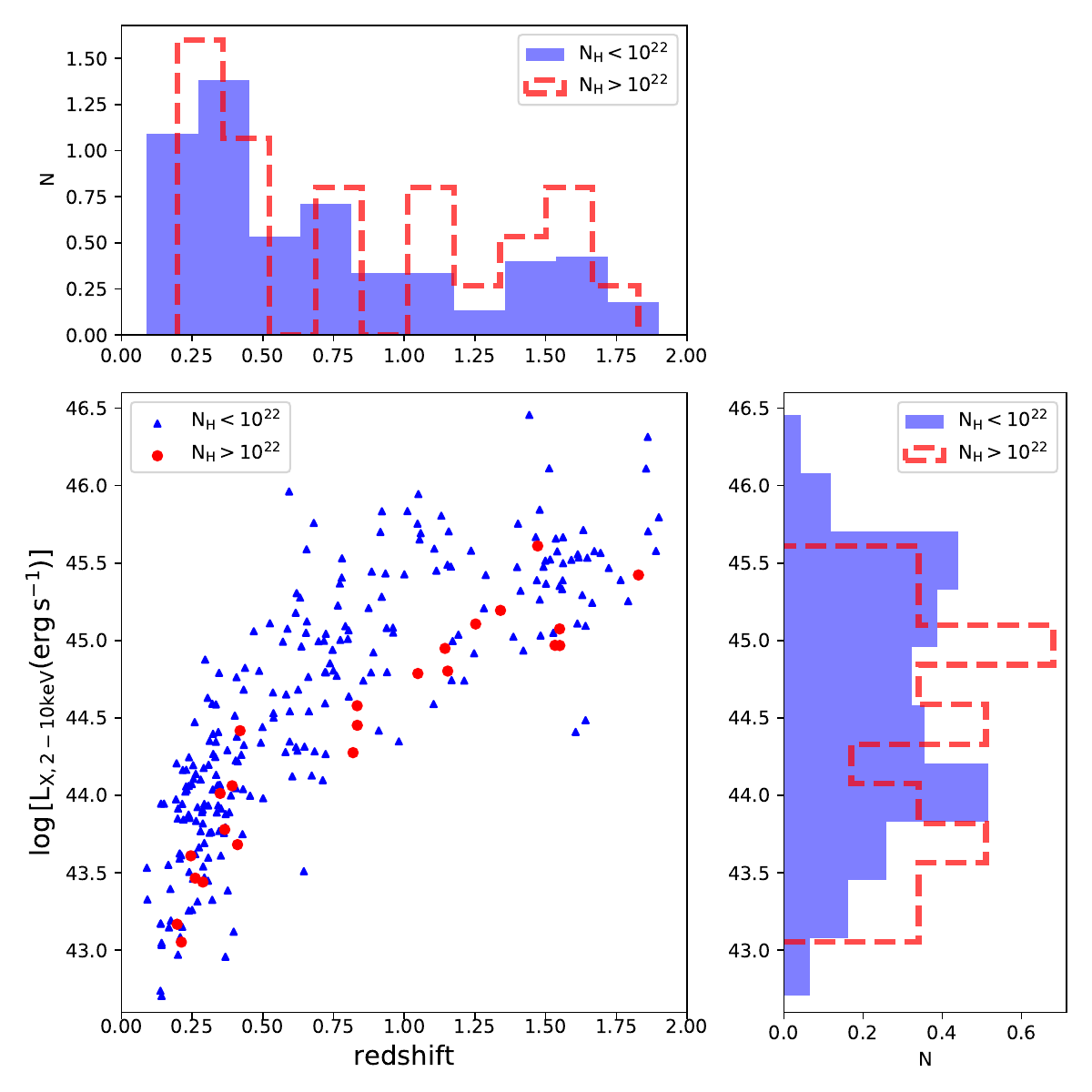}
  \caption{X-ray luminosity, L$_X$, as a function of redshift. AGN with $\rm log\,N_H >10^{22}\,cm^{-2}$ are marked in red circles, whereas sources with $\rm log\,N_H <10^{22}\,cm^{-2}$ are marked in blue triangles.}
  \label{fig_lx_redz}
\end{figure}

\begin{figure}
\centering
  \includegraphics[width=0.99\linewidth, height=7cm]{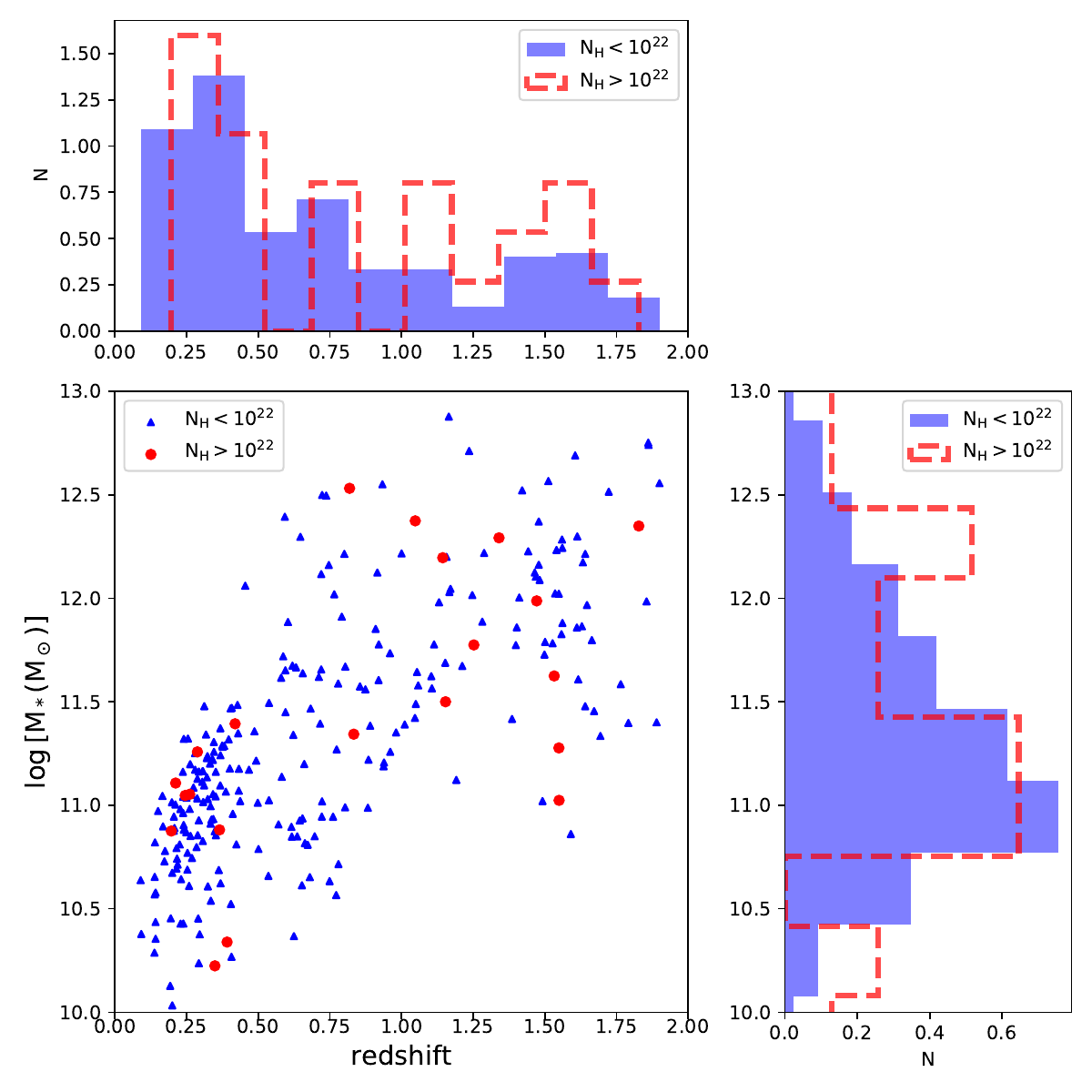}
  \caption{Stellar mass, M$_*$, as a function of redshift. AGN with $\rm log\,N_H >10^{22}\,cm^{-2}$ are marked in red circles, whereas sources with $\rm N_H <10^{22}\,cm^{-2}$ are marked in blue triangles.}
  \label{fig_mstar_redz}
\end{figure}

\begin{figure}
\centering
  \includegraphics[width=0.99\linewidth, height=7cm]{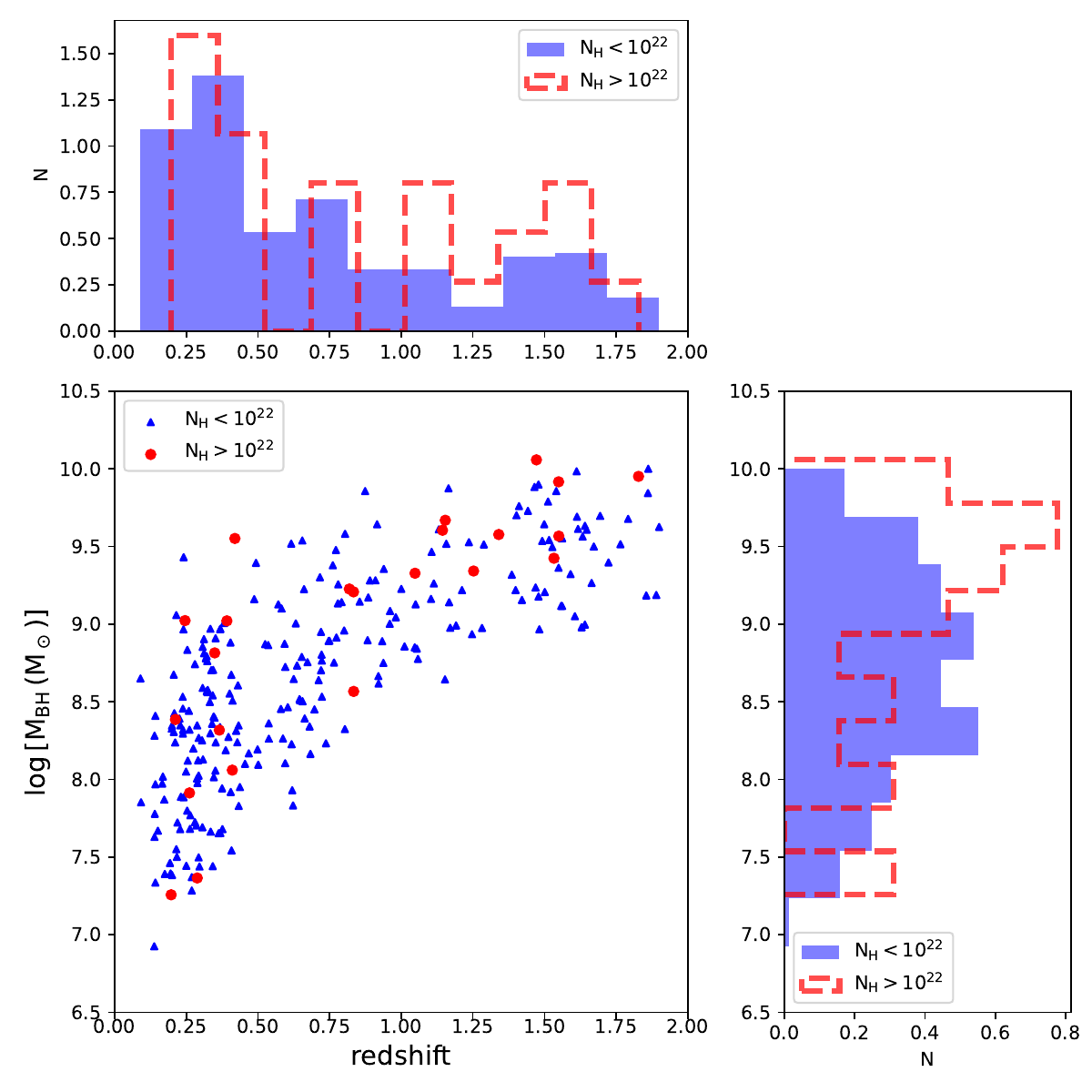}
  \caption{Black hole mass, M$_{BH}$, as a function of redshift. AGN with $\rm log\,N_H >10^{22}\,cm^{-2}$ are marked in red circles, whereas sources with $\rm N_H <10^{22}\,cm^{-2}$ are marked in blue triangles.}
  \label{fig_mbh_redz}
\end{figure} 

\subsection{Comparative analysis of SMBH characteristics, X-ray spectral attributes and host galaxy properties}
\label{sec_mbh_analysis}
In the second part of our analysis, we examined the relations among L$_{bol}$, M$_{BH}$ and M$_*$, as a function of the X-ray obscuration (N$_H$). We also investigated the correlation between the powerlaw slope, $\Gamma$, of the spectral model with the n$_{Edd}$. For that purpose, we use the 271 X-ray AGN that are common between the XMM catalogue and the \cite{Wu2022} catalogue (see Sect. \ref{sec_final_samples}). Here and in the following (sub-) sections, we characterise as AGN with high N$_H$ those sources with $\rm log\,N_H \geq 10 ^{22}cm^{-2}$, using the mode value of N$_H$ that is available in our catalogue (i.e., without using the lower and upper limits of N$_H$). There are 23 AGN with high N$_H$ values and 248 AGN with low N$_H$ values, in our dataset. The reason for using this threshold is that facilitates a better comparison with previous studies. Furthermore, we note that characterising an AGN with high N$_H$ based on requiring the lower limit of N$_H$ to be higher than $\rm 10^{23}\,cm^{-2}$ ($\rm 10^{22}\,cm^{-2}$) would result in 0 (7) sources. 

Fig. \ref{fig_lx_redz}-\ref{fig_mbh_redz}, present the distributions of L$_X$, M$_*$ and M$_{BH}$ as a function of redshift, respectively, for the 271 X-ray AGN.  All three properties increase with redshift, regardless of the $\rm N_H$ value of the source. Spearman correlation analysis reveals that there is significant correlation between L$_X$, M$_*$ and M$_{BH}$ with redshift, albeit these correlations appear higher in the case of sources with $\rm N_H < 10^{22}\,cm^{-2}$ (Table \ref{table_spearmann_basic}). We note that in this part of our analysis, only broad-line (type 1) sources are included (see Sect. \ref{sec_data}).

\begin{table}
\caption{p$-$values obtained from Spearman correlation analysis, assessing the correlations among L$_X$, M$_*$, M$_{BH}$, and redshift.}
\centering
\setlength{\tabcolsep}{1mm}
\begin{tabular}{ccc}
 \hline
 & $\rm N_H < 10^{22}\,cm^{-2}$ &  $\rm N_H > 10^{22}\,cm^{-2}$ \\
  \hline
  & p$-$value & p$-$value \\
    \hline
$\rm log\,L_{X}$-redshift & $5\times 10^{-72}$ & $3\times 10^{-12}$   \\
 \hline
$\rm log\,M_*$-redshift & $2\times 10^{-43}$ &$ 6\times 10^{-4}$  \\ 
 \hline
$\rm log\,M_{BH}$-redshift & $7\times 10^{-53}$ & $2\times 10^{-7}$ \\
  \hline
\label{table_spearmann_basic}
\end{tabular}
\end{table}

\begin{figure}
\centering
  \includegraphics[width=0.99\linewidth, height=7cm]{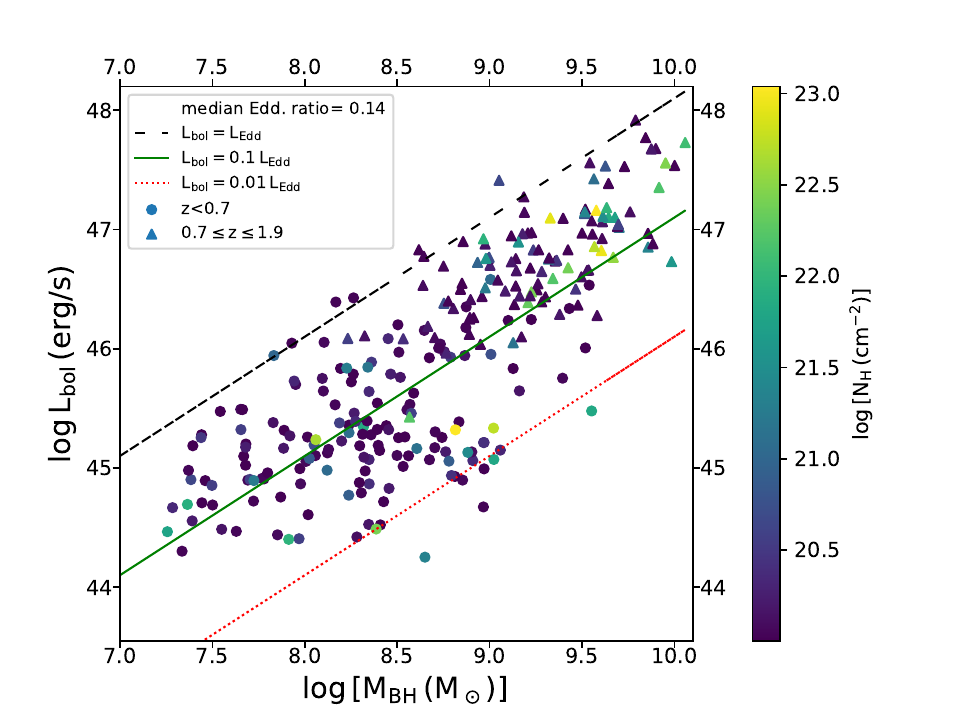}
  \caption{L$_{bol}$ as a function of M$_{BH}$. Different symbols correspond to different redshift intervals, as indicated in the legend. The lines correspond to L$_{bol}=$L$_{Edd}$ (black, dashed line), L$_{bol}=0.1$\,L$_{Edd}$ (green, black line) and L$_{bol}=0.01\,$L$_{Edd}$ (red, dotted line). Measurements are colour-coded based on the N$_H$ of each source.}
  \label{fig_lbol_mbh}
\end{figure} 

\subsubsection{L$_{bol}$ vs. $M_{BH}$}

In Fig. \ref{fig_lbol_mbh}, we plot the L$_{bol}$ as a function of M$_{BH}$. We chose to plot the L$_{bol}$ measurements that are available in the \cite{Wu2022} catalogue. We note, though, that their L$_{bol}$ calculations are in excellent agreement with the L$_{bol}$ measurements from CIGALE. Specifically, the mean difference of the two calculations is 0.04\,dex (with a dispersion of 0.36) and thus this choice does not affect our results and conclusions. The lines correspond to L$_{bol}=$\,L$_{Edd}$ (black, long-dashed line), L$_{bol}=0.1$\,L$_{Edd}$ (green, solid line) and L$_{bol}=0.01\,$L$_{Edd}$ (red, dotted line). The vast majority of our AGN lie between Eddington ratios of 0.01 to 1, with a median value of n$_{Edd}=0.14$), in agreement with previous studies \citep[e.g.,][]{Trump2009, Lusso2012, Sun2015, Suh2020, Mountrichas2023b}. Different symbols correspond to different redshift intervals, as indicated in the legend of the plot. The median value of n$_{Edd}$ is 0.09 and 0.24, at $\rm z<0.7$ and $\rm 0.7<z<1.9$, respectively.

Furthermore, measurements in Fig. \ref{fig_lbol_mbh} are colour coded based on the N$_H$ of the sources. Although the AGN sample used in this part of our analysis consists of type 1 sources, there are a few AGN that present low levels of X-ray  obscuration (N$\rm _H\sim 10^{22}\,cm^{-2}$). It is known that X-ray and optical obscuration do not necessarily coincide and cases have been reported in the literature with type 1 AGN to present X-ray obscuration \citep[e.g.,][]{Mateos2005a, Mateos2010a, Scott2011, Shimizu2018, Kamraj2019, Masoura2020}. Regarding the n$_{Edd}$ values of X-ray sources with high and low N$_H$ values, we do not find a difference for the two AGN populations. Specifically, the median values are n$_{Edd}=0.12$ and 0.15 for AGN with N$\rm _H> 10^{22}\,cm^{-2}$ and AGN with N$\rm _H< 10^{22}\,cm^{-2}$, respectively. A KS-test yields a $\rm p-value$ of 0.83 for the n$_{Edd}$ distributions of the two AGN classes.

\begin{figure}
\centering
  \includegraphics[width=0.99\linewidth, height=7cm]{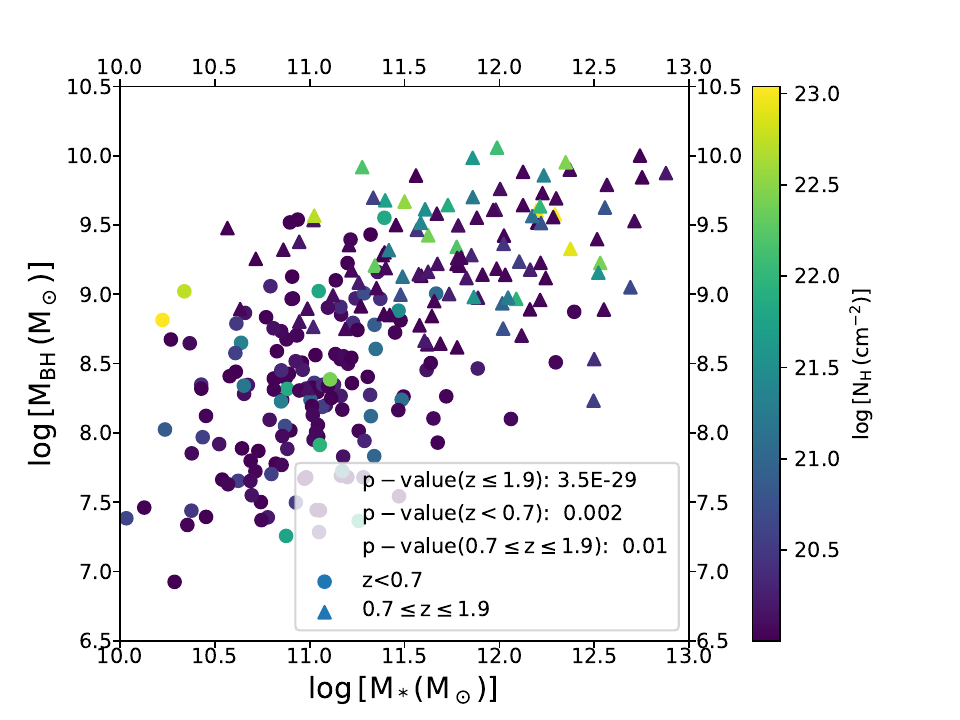}
  \caption{M$_{BH}$ as a function of M$_*$. Different symbols correspond to different redshift bins, as indicated in the legend. The $\rm p-values$  obtained by applying a Spearman correlation analysis, for the full redshift range ($\rm z<1.9)$, as well as, at low redshift ($\rm z<0.7$) and high redshift ($\rm 0.7<z<1.9$) are also shown. Measurements are colour-coded based on the N$_H$ of each source.}
  \label{fig_mbh_mstar}
\end{figure} 

\begin{figure}
\centering
  \includegraphics[width=0.99\linewidth, height=7cm]{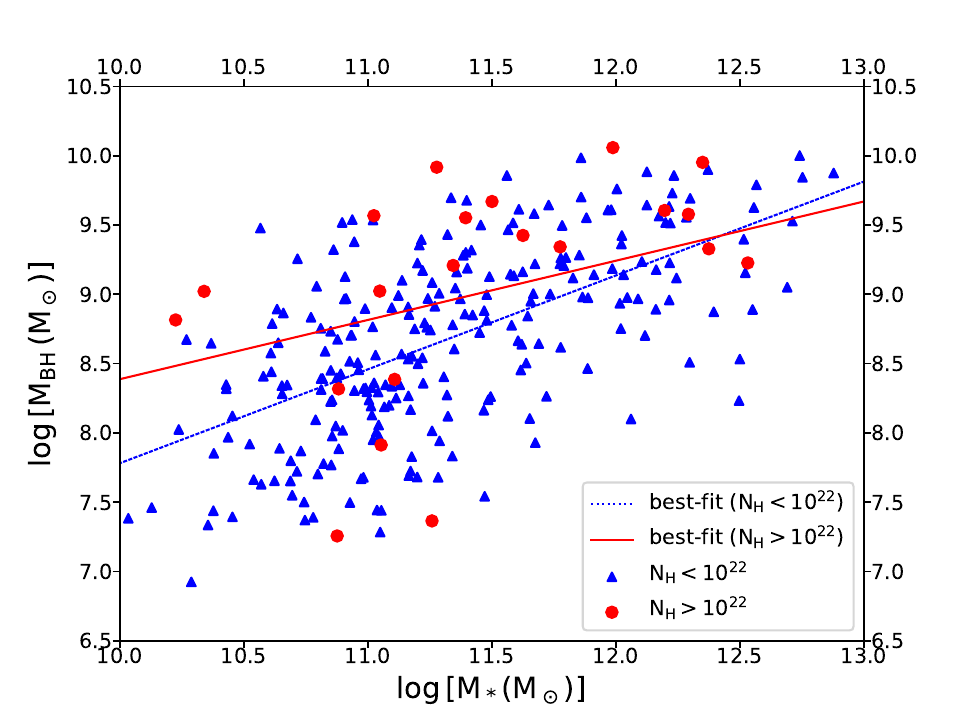}
  \caption{M$_{BH}$ as a function of M$_*$, for sources with $\rm log\,N_H>10^{22}\,cm^{-2}$ (red circles) and sources with $\rm log\,N_H<10^{22}\,cm^{-2}$ (blue triangles). Different lines represent the best fits of the different subsets, as indicated in the legend.}
  \label{fig_mbh_mstar_nh}
\end{figure}

\begin{figure}
\centering
  \includegraphics[width=0.99\linewidth, height=7cm]{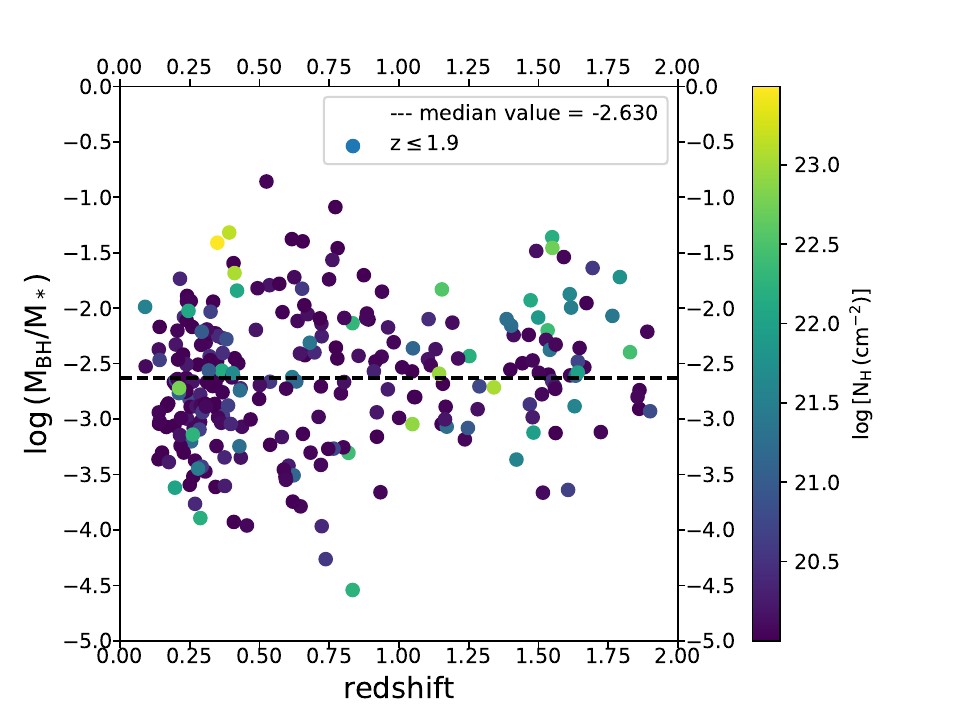}
    \includegraphics[width=0.99\linewidth, height=7cm]{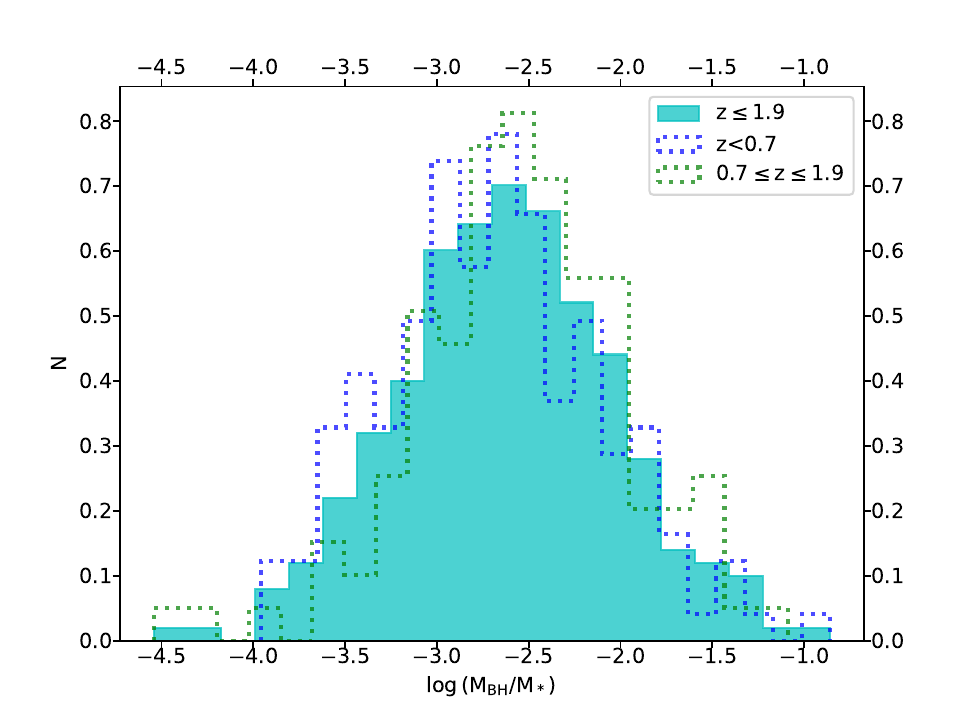}
  \caption{M$_{BH}-$M$_*$ ratio. The top panel presents the M$_{BH}-$M$_*$ ratio as a function of redshift and the bottom panel shows the distributions of the M$_{BH}-$M$_*$ ratio, at different redshift intervals.}
  \label{fig_mbh_mstar_ratio}
\end{figure} 

\begin{figure}
\centering
  \includegraphics[width=0.99\linewidth, height=7cm]{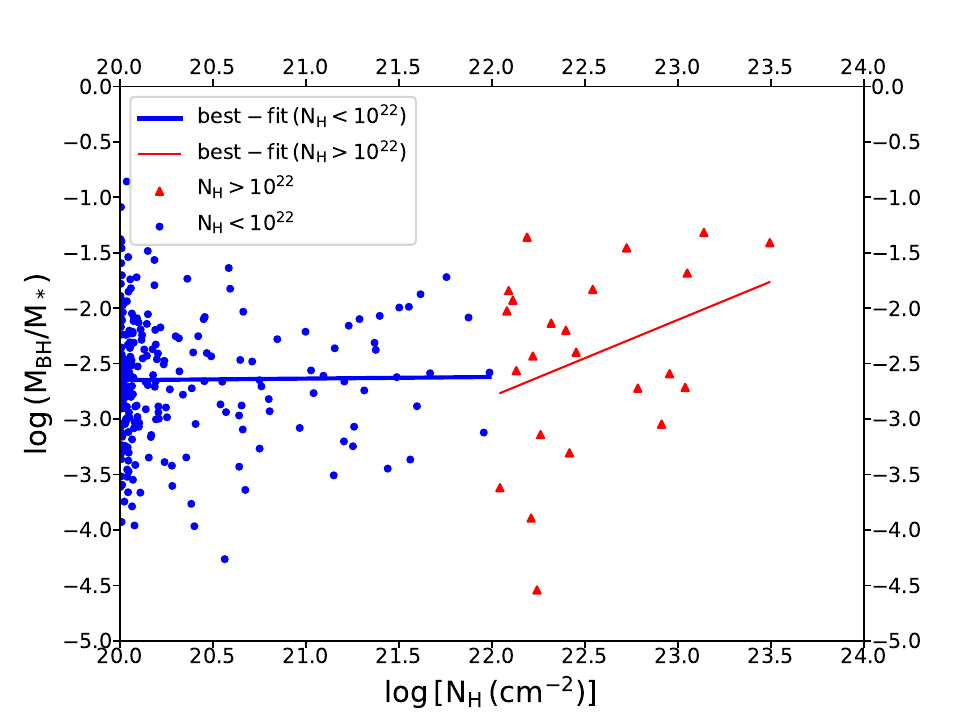}
  \caption{M$_{BH}$/M$_*$ ratio as a function of N$_H$.}
  \label{fig_mbh_mstar_ratio_nh}
\end{figure} 

\subsubsection{$M_{BH}$ versus M$_*$}

Next, we examined the correlation between  $M_{BH}$ and M$_*$. The results are presented in Fig. \ref{fig_mbh_mstar}. A strong correlation was detected between the two parameters. Applying a Spearman correlation analysis, yields a $\rm p-value$ of $3.5\times 10^{-29}$. A similar value was found using Kendall correlation analysis ($\rm p-value=8.1\times 10^{-26}$). Strong correlations between $M_{BH}$ and M$_*$ were also found when we split the dataset into two redshift intervals. The $\rm p-values$  are shown in the legend of Fig. \ref{fig_mbh_mstar}. In this case the correlations, although strong, were weaker compared to the correlation found for our full sample. This can be attributed to the narrower range of M$_{BH}$ and M$_*$ probed by the sources within distinct redshift intervals compared to the sources encompassing the entire redshift range (see Figures \ref{fig_mstar_redz} and \ref{fig_mbh_redz}). To validate this conjecture, we employed the AGN dataset spanning the complete redshift range  ($\rm z\leq 1.9$), restricted the (log of) M$_*$ and M$_{BH}$ ranges within $11.0-13.0$ and $8.5-10$, respectively (resembling the intervals embraced by AGN residing within $\rm 0.7\leq z\leq 1.9$), and carried out a Spearman correlation analysis. This procedure resulted in a $\rm p-value$ of 0.008. Similarly, we obtained a $\rm p-value$ of $6\times 10^{-4}$ by limiting the M$_*$ and M$_{BH}$ ranges to $10.0-11.5$ and $7.0-9.5$ (akin to the intervals pertinent to AGN existing within $\rm z<0.7$).

Prompted by the results of \cite{Sarria2010}, who examined three X-ray AGN at $\rm z=1-2$, with N$\rm _H=10^{22.5-23.0}\,cm^{-2}$ and observed that they were consistent with the local M$_{BH}-$M$_*$ relation, we colour coded the symbols in Fig. \ref{fig_mbh_mstar} based on the N$_H$ values of the sources. We noticed that sources with increased N$_H$ have higher M$_{BH}$ for similar M$_*$ compared to sources with lower N$_H$. This is better illustrated in Fig. \ref{fig_mbh_mstar_nh}, where we marked differently AGN with N$_H$ values above and below $10^{22}$\,cm$^{-2}$, as indicated in the legend of the plot. The different lines in Fig. \ref{fig_mbh_mstar_nh} indicate the best-fits for the various subsets, acquired through the application of least-squares analysis. As mentioned above, AGN with higher N$_H$ values tend to have more massive black holes compared to AGN with lower N$_H$ values, at similar M$_*$, at least up to $\rm log\,[M_*(M_\odot)]=12$.

We, also, examined the ratio of M$_{BH}$/M$_*$ as a function of redshift. Based on the results presented in the top panel of Fig. \ref{fig_mbh_mstar_ratio}, the M$_{BH}$/M$_*$ does not evolve with cosmic time up $\rm z<2$. The median $\rm log(M_{BH}/M_*)$ value is found at -2.63. This value is in agreement with most previous studies that measured the  M$_{BH}$/M$_*$ ratio in the local Universe \citep[e.g., -2.85][]{Haring2004}, as well as at high redshifts \citep[e.g.,][]{Suh2020, Mountrichas2023b}. The bottom panel of Fig. \ref{fig_mbh_mstar_ratio}, presents the distributions of the M$_{BH}$/M$_*$ ratio, at different redshift intervals. The median value of the M$_{BH}$/M$_*$ ratio at $\rm z<0.7$ is -2.69 and at $\rm 0.7<z<1.9$ is -2.55. Application of KS-tests confirms that the distributions are similar ($\rm p-values >0.9$, among all redshift intervals).  

The results that appear in the top panel of Fig. \ref{fig_mbh_mstar_ratio} are colour coded based on the N$_H$ of the sources. Sources with higher N$_H$ values seem to have higher M$_{BH}$/M$_*$ ratio. To examine this further, in Fig. \ref{fig_mbh_mstar_ratio_nh}, we plot the M$_{BH}$/M$_*$ ratio as a function of N$_H$. The blue line shows the best fit of the $\frac{M_{BH}}{M_*}-$log\,N$_H$ relation for log\,N$\rm _H<22\,cm^{-2}$, whereas the red line shows the best fit for  log\,N$\rm _H>22\,cm^{-2}$. The blue line is nearly flat, with a slope of 0.014 (standard error 0.079), that indicates that the  M$_{BH}$/M$_*$ ratio is almost constant for AGN with log\,N$\rm _H<22\,cm^{-2}$. Hovever, the best-fit for AGN with log\,N$\rm _H>22\,cm^{-2}$ has a slope of 0.69 (standard error 0.24), that indicates that the M$_{BH}$/M$_*$ ratio increases with N$_H$, for X-ray AGN with high N$_H$ values. Spearmann correlation analysis reveals a $\rm p-value=0.06$ for the correlation between $\frac{M_{BH}}{M_*}-$log\,N$_H$ for sources with log\,N$\rm _H>22\,cm^{-2}$, which indicates a statistical significance of $\sim 2\,\sigma$. The median value of the log\,M$_{BH}$/M$_*$ ratio is  -2.40 and -2.66 for AGN with N$_H>10^{22}$\,cm$^{-2}$ and N$_H<10^{22}$\,cm$^{-2}$, respectively. We note, though, that a larger number of sources with N$_H>10^{22}$\,cm$^{-2}$ is required to formulate robust conclusions.

In the preceding sections, we observed that obscured AGN live, on average, in galaxies with higher M$_*$ compared to unobscured. Moreover, obscured AGN tend to have higher M$_{BH}$ compared to their unobscured counterparts, at similar M$_*$.  When combined with the higher M$_{BH}$/M$_*$ ratio within the AGN population displaying elevated N$_H$ values, assuming this holds true, it implies that the SMBH growth in the obscured phase is higher than the galaxy growth. This may suggest that the growth of the M$_{BH}$ occurs first, while the early stellar mass assembly may not be so efficient \citep{Mountrichas2023b}.

\begin{figure}
\centering
  \includegraphics[width=0.99\linewidth, height=7cm]{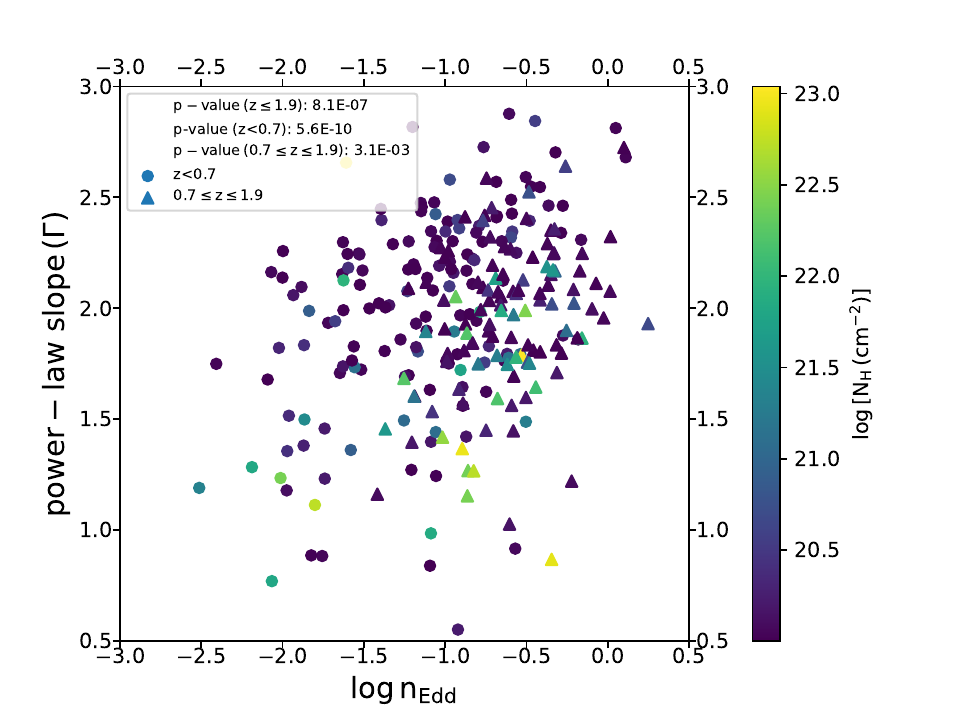}
    \includegraphics[width=0.99\linewidth, height=7cm]{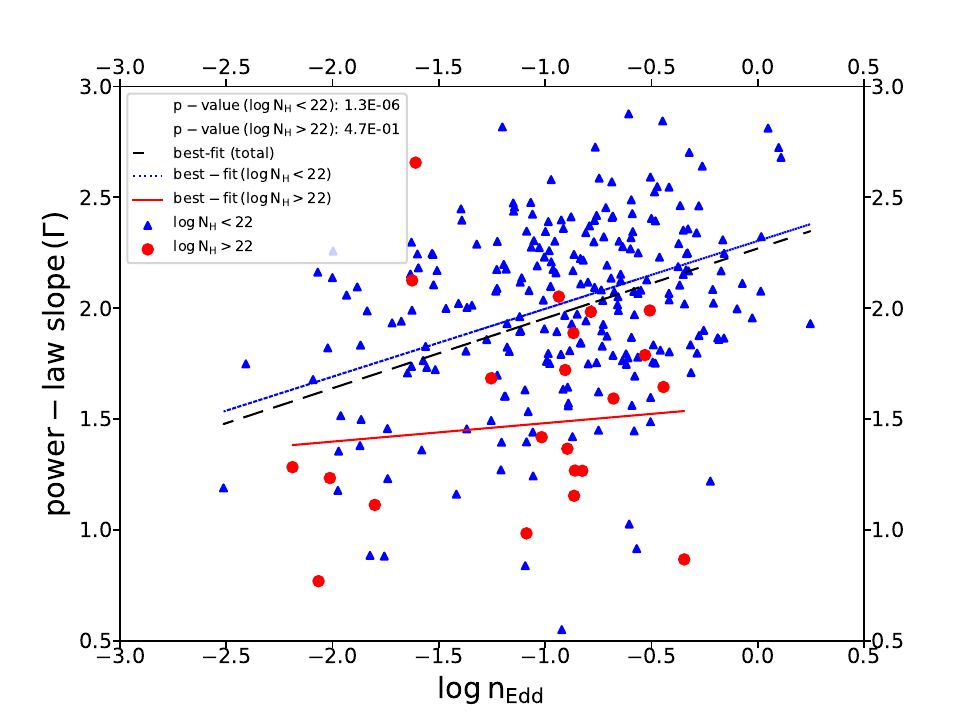}
  \caption{Spectral photon index, $\Gamma$, as a function of n$_{Edd}$. In the top panel, different symbols correspond to different redshift intervals, as indicated in the legend. The results are colour coded based on the N$_H$ values of the sources. The $\rm p-values$ obtained by applying Spearman correlation analysis, are shown in the legend. The bottom panel shows the same relation, for X-ray AGN with N$_H>10^{22}$\,cm$^{-2}$ (red circles) and AGN with N$_H<10^{22}$\,cm$^{-2}$ (blue triangles). The $\rm p-values$ for the $\Gamma-$n$_{Edd}$ correlation for each AGN population is presented in the legend of the plot. The different lines correspond to the best-fits using the total AGN sample (black, dashed line), AGN with N$_H>10^{22}$\,cm$^{-2}$ (red, solid line) and AGN with N$_H<10^{22}$\,cm$^{-2}$ (blue line).}
  \label{fig_gamma_nedd}
\end{figure} 

\begin{figure}
\centering
  \includegraphics[width=0.95\linewidth, height=7cm]{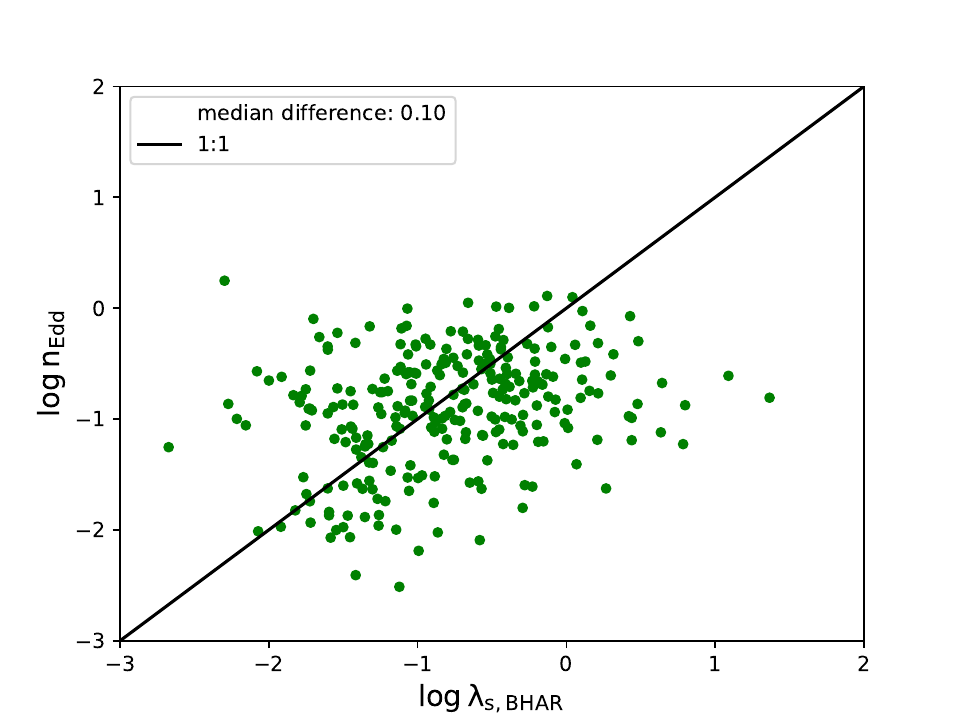}
  \caption{n$_{Edd}$ vs. $\lambda _{sBHAR}$ for the 271 AGN that are common between the XMM sample and the \cite{Wu2022} catalogue.} 
  \label{fig_lambda_vs_nedd}
\end{figure} 

\subsubsection{Powerlaw slope versus n$_{Edd}$}

We also investigated if there is a correlation between the powerlaw slope, $\Gamma$, of the X-ray spectral model and the n$_{Edd}$. The results are shown in the top panel of Fig. \ref{fig_gamma_nedd}. A strong correlation was found between $\Gamma$ and n$_{Edd}$ ($\rm p-value=8.1\times 10^{-7}$, both by applying a Spearman and a Kendall correlation analysis). Although, the correlation between the two parameters was also strong when we split the dataset into two redshift intervals,  at $\rm 0.7<z<1.9$ the $\rm p-value$ appears higher compared to the $\rm p-value$ obtained at $\rm z<0.7$ (see legend in the top panel of Fig. \ref{fig_gamma_nedd}). This is, most probable, due to the different lines used for the calculation of the M$_{BH}$ and not a result of cosmic evolution. \cite{Risaliti2009} reported a strong correlation between the two parameters when the M$_{BH}$ was calculated using the H\,$\beta$ line and a weaker correlation when the M$_{BH}$ was measured using the Mg~{\sc ii} line. There are only 12 sources in our dataset with $\rm 0.7<z<1.9$, that have M$_{BH}$ calculations from both lines, therefore we could not test, in a robust manner, this hypothesis further. We also split the dataset into high and low L$_X$, by applying a cut at $\rm log\,[L_X(ergs^{-1})]=44$. A strong correlation is found both at high and low L$_X$. Application of the Spearman correlation analysis gives $\rm $\rm p-values$=5.3\times 10^{-4}$ and $3.1\times 10^{-9}$, for $\rm log\,[L_X(ergs^{-1})]>44$ and $\rm log\,[L_X(ergs^{-1})]<44$, respectively.

We, furthermore, examined if this correlation between $\Gamma$ and n$_{Edd}$ holds for different N$_H$ values. The results are shown in the bottom panel of Fig. \ref{fig_gamma_nedd}. A strong correlation was found between $\Gamma$ and n$_{Edd}$ for AGN with N$\rm _H<10^{22}\,cm^{-2}$. Application of the Spearman correlation analysis yields $\rm p-value=1.35\times 10^{-6}$ ($\rm p-value=1.44\times 10^{-6}$, applying Kendall's correlation analysis). No correlation was found, though, for the AGN population with N$\rm _H>10^{22}\,cm^{-2}$ ($\rm p-value=0.47$ and $0.53$, from Spearman and Kendall correlation analysis, respectively). The different lines shown in the Figure, represent the best fits of the $\Gamma-$n$_{Edd}$ relation, for the total AGN sample (black, dashed line; $\Gamma = 0.315\times$\,n$_{Edd}+0.294)$, the AGN with N$\rm _H<10^{22}\,cm^{-2}$ blue line; $\Gamma = 0.306\times$\,n$_{Edd}+2.303)$) and for AGN with N$\rm _H>10^{22}\,cm^{-2}$ (red line; $\Gamma = 0.083\times$\,n$_{Edd}+1.565)$). Similar results were obtained when we split the two AGN populations in two redshift intervals, at $\rm z=0.7$. Furthermore, our observations hold when considering the associated uncertainties of $\Gamma$ and n$_{Edd}$, by utilizing the linmix module \citep{Kelly2007} that performs linear regression between two parameters, by repeatedly perturbing the datapoints within their uncertainties.

It is worth pointing out that in Fig. \ref{fig_gamma_nedd} there are some sources with non-physical values ($\Gamma >2.5$ or $\Gamma <1.5$). Although their occurrence is relatively low among AGN with N$\rm _H<10^{22}\,cm^{-2}$ ($15\%$), it becomes more pronounced ($57\%$) for AGN with higher N$_H$ values. To examine whether this might be influencing the absense of a correlation between $\Gamma$ and n$_{Edd}$ for AGN with N$\rm _H>10^{22}\,cm^{-2}$, we opted to exclude these sources and repeated the correlation analysis. However, even after this exclusion, we still did not observe a significant correlation between the two parameters for sources with elevated N$_H$ values. It's essential to note, however, that this refinement in the analysis leaves us with a considerably reduced sample size, comprising only ten AGN with N$\rm _H>10^{22}\,cm^{-2}$.

The X-ray spectral index is sensitive to the properties of the accretion disk, such as temperature and ionization state. Therefore, a correlation between the X-ray spectral index and n$_{Edd}$ for AGN with N$\rm _H<10^{22}\,cm^{-2}$ may imply that changes in the accretion rate can alter the structure and properties of the accretion disk, influencing the X-ray emission. The X-ray emission is also associated with a hot, optically thin corona of electrons above the accretion disk. The properties of the corona, such as its temperature and optical depth, can impact the X-ray spectral index. Thus, changes in the n$_{Edd}$ may lead to variations in the corona properties, affecting the X-ray spectrum. AGN with N$\rm _H<10^{22}\,cm^{-2}$ allow for a clearer view of the inner regions of the accretion disk and the corona. Furthermore, in AGN with low N$_H$, the more direct view of the accretion disk may allow for a better determination of the Eddington ratio.

Additionally, we investigated if this lack of correlation between the two parameters for the AGN population with N$\rm _H>10^{22}\,cm^{-2}$ is due to the small number of sources available. For that purpose, we randomly selected an equal number of AGN with N$\rm _H<10^{22}\,cm^{-2}$ to that with N$\rm _H>10^{22}\,cm^{-2}$ and we measured the correlation between $\Gamma$ and n$_{Edd}$. After 500 repetitions of this process, we got a median $\rm p-value$ from Spearman analysis of 0.047 (Kendall's correlation analysis yields $\rm $\rm p-value$= 0.091$). The $\rm p-value$ obtained for the simulated AGN with N$\rm _H<10^{22}\,cm^{-2}$ are about an order of magnitude lower compared to those obtained from the AGN with N$\rm _H>10^{22}\,cm^{-2}$. This indicates that the lack of correlation  between $\Gamma$ and n$_{Edd}$ for AGN with N$\rm _H>10^{22}\,cm^{-2}$ is intrinsic and not due to their smaller sample size. This result  aligns with the Comptonization models, wherein variations in the accretion rate have a more pronounced impact on the soft spectral component (which gets lost in absorbed sources) compared to the hard power-law photon index


As already note, $\lambda_{sBHAR}$ is often used as a proxy of n$_{Edd}$. \cite{Lopez2023} used X-ray selected AGN in the miniJPAS footprint and found, among others, that the n$_{Edd}$ and  $\lambda_{sBHAR}$ have a difference of 0.6\,dex. They attributed this difference to the scatter on the M$_{BH}-$M$_*$ relation of their sources. A difference between  n$_{Edd}$ and  $\lambda_{sBHAR}$ was also reported by \cite{Mountrichas2023d}, albeit lower ($\sim 0.25$\,dex), using AGN in the XMM-XXL field. Fig. \ref{fig_lambda_vs_nedd}, presents the comparison between the two parameters for the 271 X-ray AGN that are common between the XMM catalogue and the \cite{Wu2022} catalogue. We find that overall there is a good agreement between n$_{Edd}$ and  $\lambda_{sBHAR}$, with a median difference of $0.10$\,dex. We stress that this comparison includes only broad line AGN. We also confirm that this difference does not depend on N$_H$, at least up to N$_H=10^{22.5-23}$\,cm$^{-2}$, probed by our dataset. 

Since the n$_{Edd}$ measurements are in agreement with the $\lambda _{sBHAR}$ calculations, we then examined the correlation between $\Gamma$ and $\lambda _{sBHAR}$. Similar results were found with those between $\Gamma$ and n$_{Edd}$. Specifically, a strong correlation was found between the two parameters for AGN with N$\rm _H<10^{22}\,cm^{-2}$ ($\rm p-value=2.8\times 10^{-3}$), but no correlation was detected for sources with N$\rm _H>10^{22}\,cm^{-2}$ ($\rm p-value=0.95$). Prompted by these results, we utilised the larger AGN dataset used in Sect. \ref{sec_host_analysis} \citep[i.e., before matching the XMM sources with the][catalogue]{Wu2022}, to examine the correlation between $\Gamma$ and $\lambda _{sBHAR}$ for the two AGN populations. This allowed us to investigate the $\Gamma-\lambda _{sBHAR}$ relation using a significantly larger sample of obscured sources (130 AGN) that has been defined with more strict criteria (i.e., taking into account the uncertainties associated with N$_H$) and, more importantly, to include narrow-line (type 2) AGN in our investigation. Spearman correlation analysis yielded a $\rm p-value$ of $1.8\times 10^{-11}$ for AGN with N$\rm _H<10^{22}\,cm^{-2}$ and $0.31$ for AGN with N$\rm _H>10^{22}\,cm^{-2}$. We conclude that $\Gamma$ and n$_{Edd}$ are correlated at all redshift and L$_X$ probed by our dataset. However, we do not detect such correlation for AGN with increased N$_H$ values (N$\rm _H>10^{22}\,cm^{-2}$). 

Previous studies have found contradictory results regarding whether there is a correlation between $\Gamma$ and n$_{Edd}$. \cite{Shemmer2008} used 35, unabsorbed, type 1 radio quiet (RQ) AGN at $\rm z<3.2$ (25 out of the 35 sources were at $\rm z<0.5$) and found a significant correlation between $\Gamma$ and n$_{Edd}$. They concluded that a measurement of $\Gamma$ and L$_X$ can provide an estimate of n$_{Edd}$ and M$_{BH}$ with a mean uncertainty of a factor of $\leq 3$. \cite{Brightman2013} used 69 RQ AGN in the COSMOS and ECDFS fields, at $\rm z<2$ and found a significant correlation between the X-ray spectral index and n$_{Edd}$. However, based on their analysis, the scatter of the $\Gamma-$n$_{Edd}$ relation is large and thus the relation is only suitable for large samples and not for individual sources. We note, though, that the L$_{bol}$ of their sources have been calculated following a mix of different methods (SED fitting and bolometric correction to the L$_X$) that may contribute to this scatter. \cite{Trakhtenbrot2017} used 228 hard X-ray selected AGN from the Swift/Bat AGN Spectroscopic Survey (BASS) and found a very weak, but statistically significant correlation between $\Gamma$ and n$_{Edd}$. Nonetheless, their M$_{BH}$ measurements come from several different methods (see their Sect. 2.3) and 
they reported that the correlation was weaker or even absent within their different subsets. More recently, \cite{Kamraj2022} used 195 Seyfert 1 AGN in the local Universe ($\rm z<0.2$) with NuSTAR and Swift/XRT or XMM observations. They studied the correlation between the X-ray spectral index and n$_{Edd}$, using three models to fit the X-ray spectra of their sources. They found considerable scatter and no strong trend between $\Gamma$ and n$_{Edd}$ for all spectral models they applied. They concluded that the $\Gamma-$n$_{Edd}$ may not be universal or robust and could vary with the choice of the sample, the luminosity range probed and the energy range of X-ray data used in the analysis. We note, that their M$_{BH}$ estimates come from the second data release of optical measurements from the BASS survey and are inferred using a compilation of different techniques (e.g., broad-line measurements from optical spectra, stellar velocity dispersions, reverberation mapping) which could increase the scatter in the $\Gamma-$n$_{Edd}$ and weaken any possible correlation. 

Combining the results of our analysis with those from previous works, we conclude that caution has to be taken when compiling results from different studies that have used different methods to measure the M$_{BH}$ and the L$_{bol}$. Furthermore, the correlation between $\Gamma$ and n$_{Edd}$ may not be universal and could (also) depend on the level of the X-ray obscuration of the sources.


\section{Conclusions}
\label{sec_conclusions}

The parent sample of this work includes $\sim 35\,000$ X-ray AGN included in the 4XMM-DR11 catalogue, for which there are available X-ray spectra fitting measurements. To measure the host galaxy properties of these sources, we matched them with multiwavelength datasets and constructed their SEDs. Using the CIGALE SED fitting algorithm, we calculated the SFR and M$_*$ of the AGN. Our analysis demanded stringent selection criteria to ensure that only AGN with dependable X-ray spectral and SED fitting measurements were incorporated in our investigation.  There were 1\,443 AGN that fulfilled these requirements. We, then, applied strict criteria using the N$_H$ values and their associated uncertainties to classify sources into obscured and unobscured and compare their SMBH and host galaxy properties. Our main findings are the following: 

\begin{itemize}

\item[$\bullet$] Obscured AGN tend to live in more massive systems (by $\sim 0.1$\,dex) compared to unobscured. The difference, although small, appears to be statistically significant. Obscured sources also tend to live in galaxies with lower SFR (by $\sim 0.25$\,dex) compared to their unobscured counterparts, however, this different is not statistically significant. The results do not depend on the N$_H$ threshold used to classify AGN (N$\rm _H=10^{23}\,cm^{-2}$ or N$\rm _H=10^{22}\,cm^{-2}$), but the differences are not statistically significant at high L$_X$ ($\rm log\,[L_{X}(ergs^{-1})]>44$) or if we disregard the errors associated with the N$_H$ measurements. 

\item[$\bullet$] Unobscured AGN have, on average, higher specific black hole accretion rates (a proxy of the Eddington ratio) compared to unobscured sources. The difference is $0.1-0.2$\,dex and appears to have a high statistical significance.

\end{itemize}

Furthermore, we cross-matched our dataset with the catalogue of \cite{Wu2022} that includes measurements for the M$_{BH}$, L$_{bol}$ and Eddington ratio, among others, for $\sim 750,000$ QSOs from SDSS. There are 271 type 1 AGN common in the two datasets, up to redshift of 1.9. We studied the n$_{Edd}$ and M$_{BH}$/M$_*$ ratio of the two AGN populations and we examined if there is a correlation between $\Gamma$ and n$_{Edd}$. Our main results are summarized as follows:

\begin{itemize}

\item[$\bullet$] Type 1 AGN with N$\rm _H>10^{22}\,cm^{-2}$ tend to have higher M$_{BH}$ compared to type 1 sources with lower N$_H$ values, at similar M$_*$.

\item[$\bullet$] For type 1 AGN, the M$_{BH}$/M$_*$ ratio is nearly constant with N$_H$ up to N$_H=10^{22}$\,cm$^{-2}$. However, our results suggest that the M$_{BH}$/M$_*$ ratio increases at higher N$_H$ values.

\item[$\bullet$] A correlation is found between the spectral photon index, $\Gamma$, and the Eddington ratio, n$_{Edd}$, for type 1 AGN with N$_H<10^{22}$\,cm$^{-2}$.

\end{itemize}

The findings of our study indicate that the inconsistent results among previous studies regarding the host galaxy properties of obscured and unobscured AGN could be, mainly, due to the different luminosities probed. Additionally, our analysis indicates that during the obscured phase, SMBH growth exceeds that of the host galaxies, potentially implying a sequential growth pattern where M$_{BH}$ develops first, while early stellar mass assembly might not be as efficient. Furthermore, the disparities in previous research regarding the correlation between $\Gamma$ and n$_{Edd}$ may also be partly attributed to differences in the fraction of obscured and unobscured AGN within the samples used in these previous studies.

\begin{acknowledgements}
This project has received funding from the European Union's Horizon 2020 research and innovation program under grant agreement no. 101004168, the XMM2ATHENA project.
This research has made use of data obtained from the 4XMM XMM-Newton serendipitous source catalogue compiled by the 10 institutes of the XMM-Newton Survey Science Centre selected by ESA.

Funding for the Sloan Digital Sky Survey V has been provided by the Alfred P. Sloan Foundation, the Heising-Simons Foundation, the National Science Foundation, and the Participating Institutions. SDSS acknowledges support and resources from the Center for High-Performance Computing at the University of Utah. The SDSS web site is \url{www.sdss.org}. SDSS is managed by the Astrophysical Research Consortium for the Participating Institutions of the SDSS Collaboration, including the Carnegie Institution for Science, Chilean National Time Allocation Committee (CNTAC) ratified researchers, the Gotham Participation Group, Harvard University, Heidelberg University, The Johns Hopkins University, L Ecole polytechnique federale de Lausanne (EPFL), Leibniz-Institut fur Astrophysik Potsdam (AIP), Max-Planck-Institut fur Astronomie (MPIA Heidelberg), Max-Planck-Institut fur Extraterrestrische Physik (MPE), Nanjing University, National Astronomical Observatories of China (NAOC), New Mexico State University, The Ohio State University, Pennsylvania State University, Smithsonian Astrophysical Observatory, Space Telescope Science Institute (STScI), the Stellar Astrophysics Participation Group, Universidad Nacional Aut\'{o}noma de M\'{e}xico, University of Arizona, University of Colorado Boulder, University of Illinois at Urbana-Champaign, University of Toronto, University of Utah, University of Virginia, Yale University, and Yunnan University.

The Pan-STARRS1 Surveys (PS1) and the PS1 public science archive have been made possible through contributions by the Institute for Astronomy, the University of Hawaii, the Pan-STARRS Project Office, the Max-Planck Society and its participating institutes, the Max Planck Institute for Astronomy, Heidelberg and the Max Planck Institute for Extraterrestrial Physics, Garching, The Johns Hopkins University, Durham University, the University of Edinburgh, the Queen's University Belfast, the Harvard-Smithsonian Center for Astrophysics, the Las Cumbres Observatory Global Telescope Network Incorporated, the National Central University of Taiwan, the Space Telescope Science Institute, the National Aeronautics and Space Administration under Grant No. NNX08AR22G issued through the Planetary Science Division of the NASA Science Mission Directorate, the National Science Foundation Grant No. AST-1238877, the University of Maryland, Eotvos Lorand University (ELTE), the Los Alamos National Laboratory, and the Gordon and Betty Moore Foundation.

This publication makes use of data products from the Wide-field Infrared Survey Explorer, which is a joint project of the University of California, Los Angeles, and the Jet Propulsion Laboratory/California Institute of Technology, funded by the National Aeronautics and Space Administration.
This research has made use of TOPCAT version 4.8 \citep{Taylor2005} and Astropy \citep{Astropy2022}.

\end{acknowledgements}

\bibliography{mybib}

\begin{thebibliography}{84}
\expandafter\ifx\csname natexlab\endcsname\relax\def\natexlab#1{#1}\fi

\bibitem[{Aird {et~al.}(2018)Aird, Coil, \& Georgakakis}]{Aird2018}
Aird, J., Coil, A.~L., \& Georgakakis, A. 2018, Monthly Notices of the Royal
  Astronomical Society, 474, 1225

\bibitem[{Arnaud(1996)}]{Arnaud1996}
Arnaud, K.~A. 1996, ASPC, 101, 17

\bibitem[{Bolton {et~al.}(2012)Bolton, Schlegel, Aubourg, Bailey, Bhardwaj,
  Brownstein, Burles, Chen, Dawson, Eisenstein, Gunn, Knapp, Loomis, Lupton,
  Maraston, Muna, Myers, Olmstead, Padmanabhan, P{\^{a}}ris, Percival,
  Petitjean, Rockosi, Ross, Schneider, Shu, Strauss, Thomas, Tremonti, Wake,
  Weaver, \& Wood-Vasey}]{Bolton2012}
Bolton, A.~S., Schlegel, D.~J., Aubourg, {\'{E}}., {et~al.} 2012, The
  Astronomical Journal, 144, 144

\bibitem[{Boquien {et~al.}(2019)Boquien, Burgarella, Roehlly, Buat, Ciesla,
  Corre, Inoue, \& Salas}]{Boquien2019}
Boquien, M., Burgarella, D., Roehlly, Y., {et~al.} 2019, Astronomy {\&}
  Astrophysics, 622, A103

\bibitem[{Boyle {et~al.}(2000)Boyle, Shanks, Croom, Smith, Miller, Loaring, \&
  Heymans}]{Boyle2000}
Boyle, B.~J., Shanks, T., Croom, S.~M., {et~al.} 2000, Monthly Notices of the
  Royal Astronomical Society, 317, 1014

\bibitem[{Brightman {et~al.}(2013)Brightman, Silverman, Mainieri, Ueda,
  Schramm, Matsuoka, Nagao, Steinhardt, Kartaltepe, Sanders, Treister, Shemmer,
  Brandt, Brusa, Comastri, Ho, Lanzuisi, Lusso, Nandra, Salvato, Zamorani,
  Akiyama, Alexander, Bongiorno, Capak, Civano, Moro, Doi, Elvis, Hasinger,
  Laird, Masters, Mignoli, Ohta, Schawinski, \& Taniguchi}]{Brightman2013}
Brightman, M., Silverman, J.~D., Mainieri, V., {et~al.} 2013, Monthly Notices
  of the Royal Astronomical Society, 433, 2485

\bibitem[{Bruzual \& Charlot(2003)}]{Bruzual_Charlot2003}
Bruzual, G. \& Charlot, S. 2003, MNRAS, 344, 1000

\bibitem[{Buat {et~al.}(2019)Buat, Ciesla, Boquien, Ma{\l}ek, \&
  Burgarella}]{Buat2019}
Buat, V., Ciesla, L., Boquien, M., Ma{\l}ek, K., \& Burgarella, D. 2019,
  Astronomy {\&} Astrophysics, 632, A79

\bibitem[{Buat {et~al.}(2021)Buat, Mountrichas, Yang, Boquien, Roehlly,
  Burgarella, Stalevski, Ciesla, \& Theul{\'{e}}}]{Buat2021}
Buat, V., Mountrichas, G., Yang, G., {et~al.} 2021, A\&A, 654, A93

\bibitem[{Buchner(2019)}]{Buchner2019}
Buchner, J. 2019, Publications of the Astronomical Society of the Pacific, 131,
  108005

\bibitem[{Buchner {et~al.}(2021)Buchner, Brightman, Balokovic, Wada, Bauer, \&
  Nandra}]{Buchner2021}
Buchner, J., Brightman, M., Balokovic, M., {et~al.} 2021, Astronomy \&
  Astrophysics, 651, A58

\bibitem[{{Buchner} {et~al.}(2014)}]{Buchner2014}
{Buchner}, J. {et~al.} 2014, A\&A, 564, 125

\bibitem[{Charlot \& Fall(2000)}]{Charlot_Fall_2000}
Charlot, S. \& Fall, S.~M. 2000, ApJ, 539, 718

\bibitem[{Ciotti \& Ostriker(1997)}]{Ciotti1997}
Ciotti, L. \& Ostriker, J.~P. 1997, The Astrophysical Journal, 487, L105

\bibitem[{Collaboration {et~al.}(2022)Collaboration, Price-Whelan, Lim, Earl,
  Starkman, Bradley, Shupe, Patil, Corrales, Brasseur, Nöthe, Donath,
  Tollerud, Morris, Ginsburg, Vaher, Weaver, Tocknell, Jamieson, van Kerkwijk,
  Robitaille, Merry, Bachetti, Günther, Aldcroft, Alvarado-Montes, Archibald,
  Bódi, Bapat, Barentsen, Bazán, Biswas, Boquien, Burke, Cara, Cara, Conroy,
  Conseil, Craig, Cross, Cruz, D'Eugenio, Dencheva, Devillepoix, Dietrich,
  Eigenbrot, Erben, Ferreira, Foreman-Mackey, Fox, Freij, Garg, Geda, Glattly,
  Gondhalekar, Gordon, Grant, Greenfield, Groener, Guest, Gurovich, Handberg,
  Hart, Hatfield-Dodds, Homeier, Hosseinzadeh, Jenness, Jones, Joseph,
  Kalmbach, Karamehmetoglu, Kałuszyński, Kelley, Kern, Kerzendorf, Koch,
  Kulumani, Lee, Ly, Ma, MacBride, Maljaars, Muna, Murphy, Norman, O'Steen,
  Oman, Pacifici, Pascual, Pascual-Granado, Patil, Perren, Pickering, Rastogi,
  Roulston, Ryan, Rykoff, Sabater, Sakurikar, Salgado, Sanghi, Saunders,
  Savchenko, Schwardt, Seifert-Eckert, Shih, Jain, Shukla, Sick, Simpson,
  Singanamalla, Singer, Singhal, Sinha, Sipőcz, Spitler, Stansby, Streicher,
  Šumak, Swinbank, Taranu, Tewary, Tremblay, de~Val-Borro, Kooten, Vasović,
  Verma, de~Miranda~Cardoso, Williams, Wilson, Winkel, Wood-Vasey, Xue,
  Yoachim, ZHANG, \& Zonca}]{Astropy2022}
Collaboration, T.~A., Price-Whelan, A.~M., Lim, P.~L., {et~al.} 2022, ApJ
  [\eprint[arXiv]{2206.14220}]

\bibitem[{{Dale} {et~al.}(2014){Dale}, {Helou}, {Magdis}, {Armus},
  {D{\'{\i}}az-Santos}, \& {Shi}}]{Dale2014}
{Dale}, D.~A., {Helou}, G., {Magdis}, G.~E., {et~al.} 2014, ApJ, 784, 83

\bibitem[{Davis \& Laor(2011)}]{Davis2011}
Davis, S.~W. \& Laor, A. 2011, The Astrophysical Journal, 728, 98

\bibitem[{Esparza-Arredondo {et~al.}(2021)Esparza-Arredondo, Gonzalez-Martín,
  Dultzin, Masegosa, Ramos-Almeida, García-Bernete, Fritz, \&
  Osorio-Clavijo}]{Esparza_Arredondo_2021}
Esparza-Arredondo, D., Gonzalez-Martín, O., Dultzin, D., {et~al.} 2021,
  Astronomy \& Astrophysics, 651, A91

\bibitem[{Evans {et~al.}(2020)Evans, Page, Osborne, Beardmore, Willingale,
  Burrows, Kennea, Perri, Capalbi, Tagliaferri, \& Cenko}]{Evans2020}
Evans, P.~A., Page, K.~L., Osborne, J.~P., {et~al.} 2020, The Astrophysical
  Journal Supplement Series, 247, 54

\bibitem[{{Ferrarese} \& {Merritt}(2000)}]{Ferrarese2000}
{Ferrarese}, L. \& {Merritt}, D. 2000, ApJ, 539, 9

\bibitem[{Georgakakis {et~al.}(2017)Georgakakis, Aird, Schulze, Dwelly,
  Salvato, Nandra, Merloni, \& Schneider}]{Georgakakis2017}
Georgakakis, A., Aird, J., Schulze, A., {et~al.} 2017, MNRAS, 471, 1976

\bibitem[{Georgantopoulos {et~al.}(2023)Georgantopoulos, Pouliasis,
  Mountrichas, der Wel, Marchesi, \& Lanzuisi}]{Georgantopoulos2023}
Georgantopoulos, I., Pouliasis, E., Mountrichas, G., {et~al.} 2023, Astronomy
  \& Astrophysics, 673, A67

\bibitem[{{H{\"a}ring} \& {Rix}(2004)}]{Haring2004}
{H{\"a}ring}, N. \& {Rix}, H.-W. 2004, ApJl, 604, L89

\bibitem[{Hopkins {et~al.}(2006)Hopkins, Hernquist, Cox, Matteo, Robertson, \&
  Springel}]{Hopkins2006}
Hopkins, P.~F., Hernquist, L., Cox, T.~J., {et~al.} 2006, The Astrophysical
  Journal Supplement Series, 163, 1

\bibitem[{Kamraj {et~al.}(2019)Kamraj, Baloković, Brightman, Stern, Harrison,
  Assef, Koss, Oh, \& Walton}]{Kamraj2019}
Kamraj, N., Baloković, M., Brightman, M., {et~al.} 2019, The Astrophysical
  Journal, 887, 255

\bibitem[{Kamraj {et~al.}(2022)Kamraj, Brightman, Harrison, Stern,
  Garc{\'{\i}}a, Balokovi{\'{c}}, Ricci, Koss, Mej{\'{\i}}a-Restrepo, Oh,
  Powell, \& Urry}]{Kamraj2022}
Kamraj, N., Brightman, M., Harrison, F.~A., {et~al.} 2022, The Astrophysical
  Journal, 927, 42

\bibitem[{Kelly(2007)}]{Kelly2007}
Kelly, B.~C. 2007, The Astrophysical Journal, 665, 1489

\bibitem[{Koutoulidis {et~al.}(2022)Koutoulidis, Mountrichas, Georgantopoulos,
  Pouliasis, \& Plionis}]{Koutoulidis2022}
Koutoulidis, L., Mountrichas, G., Georgantopoulos, I., Pouliasis, E., \&
  Plionis, M. 2022, Astronomy {\&} Astrophysics, 658, A35

\bibitem[{{Lanzuisi} {et~al.}(2017)}]{Lanzuisi2017}
{Lanzuisi}, G. {et~al.} 2017, A\&A, 602, 13

\bibitem[{Lopez {et~al.}(2023)Lopez, Brusa, Bonoli, Shankar, Acharya, Laloux,
  Dolag, Georgakakis, \& Lapi}]{Lopez2023}
Lopez, I.~E., Brusa, M., Bonoli, S., {et~al.} 2023, Astronomy {\&}
  Astrophysics, 672, A137

\bibitem[{{Lusso} {et~al.}(2012)}]{Lusso2012}
{Lusso}, E. {et~al.} 2012, MNRAS, 425, 623

\bibitem[{Lyke {et~al.}(2020)Lyke, Higley, McLane, Schurhammer, Myers, Ross,
  Dawson, Chabanier, Martini, Busca, du~Mas~des Bourboux, Salvato,
  Streblyanska, Zarrouk, Burtin, Anderson, Bautista, Bizyaev, Brandt,
  Brinkmann, Brownstein, Comparat, Green, de~la Macorra, Guti{\'{e}}rrez, Hou,
  Newman, Palanque-Delabrouille, P{\^{a}}ris, Percival, Petitjean, Rich, Rossi,
  Schneider, Smith, Vivek, \& Weaver}]{Lyke2020}
Lyke, B.~W., Higley, A.~N., McLane, J.~N., {et~al.} 2020, The Astrophysical
  Journal Supplement Series, 250, 8

\bibitem[{Magorrian {et~al.}(1998)}]{Magorrian1998}
Magorrian, J. {et~al.} 1998, AJ, 115, 2285

\bibitem[{Ma{\l}ek {et~al.}(2018)Ma{\l}ek, Buat, Roehlly, Burgarella, Hurley,
  Shirley, Duncan, Efstathiou, Papadopoulos, Vaccari, Farrah, Marchetti, \&
  Oliver}]{Malek2018}
Ma{\l}ek, K., Buat, V., Roehlly, Y., {et~al.} 2018, Astronomy {\&}
  Astrophysics, 620, A50

\bibitem[{Marinucci {et~al.}(2016)Marinucci, Bianchi, Matt, Alexander,
  Baloković, Bauer, Brandt, Gandhi, Guainazzi, Harrison, Iwasawa, Koss,
  Madsen, Nicastro, Puccetti, Ricci, Stern, \& Walton}]{Marinucci2016}
Marinucci, A., Bianchi, S., Matt, G., {et~al.} 2016, Monthly Notices of the
  Royal Astronomical Society: Letters, 456, L94–L98

\bibitem[{Masoura {et~al.}(2020)Masoura, Georgantopoulos, Mountrichas, Vignali,
  Koulouridis, Chiappetti, Fotopoulou, Paltani, \& Pierre}]{Masoura2020}
Masoura, V.~A., Georgantopoulos, I., Mountrichas, G., {et~al.} 2020, Astronomy
  {\&} Astrophysics, 638, A45

\bibitem[{Masoura {et~al.}(2021)Masoura, Mountrichas, Georgantopoulos, \&
  Plionis}]{Masoura2021}
Masoura, V.~A., Mountrichas, G., Georgantopoulos, I., \& Plionis, M. 2021,
  Astronomy {\&} Astrophysics, 646, A167

\bibitem[{Masoura {et~al.}(2018)Masoura, Mountrichas, Georgantopoulos, Ruiz,
  Magdis, \& Plionis}]{Masoura2018}
Masoura, V.~A., Mountrichas, G., Georgantopoulos, I., {et~al.} 2018, A\&A, 618,
  31

\bibitem[{Mateos {et~al.}(2005)Mateos, Barcons, Carrera, Ceballos, Hasinger,
  Lehmann, Fabian, \& Streblyanska}]{Mateos2005a}
Mateos, S., Barcons, X., Carrera, F.~J., {et~al.} 2005, Astronomy \&
  Astrophysics, 444, 79

\bibitem[{Mateos {et~al.}(2010)Mateos, Carrera, Page, Watson, Corral, Tedds,
  Ebrero, Krumpe, Schwope, \& Ceballos}]{Mateos2010a}
Mateos, S., Carrera, F.~J., Page, M.~J., {et~al.} 2010, Astronomy and
  Astrophysics, 510, A35

\bibitem[{Merloni {et~al.}(2014)Merloni, Bongiorno, Brusa, Iwasawa, Mainieri,
  Magnelli, Salvato, Berta, Cappelluti, Comastri, Fiore, Gilli, \&
  Koekemoer}]{Merloni2014}
Merloni, A., Bongiorno, A., Brusa, M., {et~al.} 2014, Monthly Notices of the
  Royal Astronomical Society, 437, 3550

\bibitem[{Mountrichas \& Buat(2023)}]{Mountrichas2023d}
Mountrichas, G. \& Buat, V. 2023, Astronomy \& Astrophysics, 679, A151

\bibitem[{Mountrichas {et~al.}(2021{\natexlab{a}})Mountrichas, Buat,
  Georgantopoulos, Yang, Masoura, Boquien, \& Burgarella}]{Mountrichas2021b}
Mountrichas, G., Buat, V., Georgantopoulos, I., {et~al.} 2021{\natexlab{a}},
  Astronomy {\&} Astrophysics, 653, A70

\bibitem[{Mountrichas {et~al.}(2021{\natexlab{b}})Mountrichas, Buat, Yang,
  Boquien, Burgarella, \& Ciesla}]{Mountrichas2021a}
Mountrichas, G., Buat, V., Yang, G., {et~al.} 2021{\natexlab{b}}, Astronomy
  {\&} Astrophysics, 646, A29

\bibitem[{Mountrichas {et~al.}(2021{\natexlab{c}})Mountrichas, Buat, Yang,
  Boquien, Burgarella, Ciesla, Malek, \& Shirley}]{Mountrichas2021c}
Mountrichas, G., Buat, V., Yang, G., {et~al.} 2021{\natexlab{c}}, Astronomy
  {\&} Astrophysics, 653, A74

\bibitem[{Mountrichas {et~al.}(2022{\natexlab{a}})Mountrichas, Buat, Yang,
  Boquien, Burgarella, Ciesla, Malek, \& Shirley}]{Mountrichas2022b}
Mountrichas, G., Buat, V., Yang, G., {et~al.} 2022{\natexlab{a}}, Astronomy
  {\&} Astrophysics, 663, A130

\bibitem[{Mountrichas {et~al.}(2022{\natexlab{b}})Mountrichas, Buat, Yang,
  Boquien, Ni, Pouliasis, Burgarella, Theule, \&
  Georgantopoulos}]{Mountrichas2022c}
Mountrichas, G., Buat, V., Yang, G., {et~al.} 2022{\natexlab{b}}, Astronomy
  {\&} Astrophysics, 667, A145

\bibitem[{Mountrichas {et~al.}(2019)Mountrichas, Georgakakis, \&
  Georgantopoulos}]{Mountrichas2019}
Mountrichas, G., Georgakakis, A., \& Georgantopoulos, I. 2019, Monthly Notices
  of the Royal Astronomical Society, 483, 1374

\bibitem[{Mountrichas {et~al.}(2022{\natexlab{c}})Mountrichas, Masoura,
  Xilouris, Georgantopoulos, Buat, \& Paspaliaris}]{Mountrichas2022a}
Mountrichas, G., Masoura, V.~A., Xilouris, E.~M., {et~al.} 2022{\natexlab{c}},
  Astronomy {\&} Astrophysics, 661, A108

\bibitem[{Mountrichas \& Shankar(2023)}]{Mountrichas2023}
Mountrichas, G. \& Shankar, F. 2023, Monthly Notices of the Royal Astronomical
  Society, 518, 2088

\bibitem[{Mountrichas {et~al.}(2023{\natexlab{a}})Mountrichas, Yang, Buat,
  Darvish, Boquien, Ni, Burgarella, \& Ciesla}]{Mountrichas2023c}
Mountrichas, G., Yang, G., Buat, V., {et~al.} 2023{\natexlab{a}}, Astronomy \&
  Astrophysics, 675, A137

\bibitem[{Mountrichas {et~al.}(2023{\natexlab{b}})Mountrichas, Yang, Buat,
  Darvish, Boquien, Ni, Burgarella, \& Ciesla}]{Mountrichas2023b}
Mountrichas, G., Yang, G., Buat, V., {et~al.} 2023{\natexlab{b}}, The relation
  of cosmic environment and morphology with the star formation and stellar
  populations of AGN and non-AGN galaxies

\bibitem[{Nenkova {et~al.}(2002)Nenkova, Ivezi{\'{c}}, \&
  Elitzur}]{Nenkova2002}
Nenkova, M., Ivezi{\'{c}}, {\v{Z}}., \& Elitzur, M. 2002, The Astrophysical
  Journal, 570, L9

\bibitem[{Netzer(2015)}]{Netzer2015}
Netzer, H. 2015, Annual Review of Astronomy and Astrophysics, 53, 365

\bibitem[{Ogawa {et~al.}(2021)Ogawa, Ueda, Tanimoto, \& Yamada}]{Ogawa2021}
Ogawa, S., Ueda, Y., Tanimoto, A., \& Yamada, S. 2021, The Astrophysical
  Journal, 906, 84

\bibitem[{Pouliasis {et~al.}(2022)Pouliasis, Mountrichas, Georgantopoulos,
  Ruiz, Gilli, Koulouridis, Akiyama, Ueda, Garrel, Nagao, Paltani, Pierre,
  Toba, \& Vignali}]{Pouliasis2022}
Pouliasis, E., Mountrichas, G., Georgantopoulos, I., {et~al.} 2022, Astronomy
  {\&} Astrophysics, 667, A56

\bibitem[{Ricci {et~al.}(2018)Ricci, Ho, Fabian, Trakhtenbrot, Koss, Ueda,
  Lohfink, Shimizu, Bauer, Mushotzky, Schawinski, Paltani, Lamperti, Treister,
  \& Oh}]{Ricci2018}
Ricci, C., Ho, L.~C., Fabian, A.~C., {et~al.} 2018, Monthly Notices of the
  Royal Astronomical Society, 480, 1819

\bibitem[{Risaliti {et~al.}(2009)Risaliti, Young, \& Elvis}]{Risaliti2009}
Risaliti, G., Young, M., \& Elvis, M. 2009, The Astrophysical Journal, 700, L6

\bibitem[{Ruiz {et~al.}(2018)Ruiz, Corral, Mountrichas, \&
  Georgantopoulos}]{Ruiz2018}
Ruiz, A., Corral, A., Mountrichas, G., \& Georgantopoulos, I. 2018, Astronomy
  {\&} Astrophysics, 618, A52

\bibitem[{Sarria {et~al.}(2010)Sarria, Maiolino, Franca, Pozzi, Fiore, Marconi,
  Vignali, \& Comastri}]{Sarria2010}
Sarria, J.~E., Maiolino, R., Franca, F.~L., {et~al.} 2010, Astronomy \&
  Astrophysics, 522, L3

\bibitem[{Scott {et~al.}(2011)Scott, Stewart, Mateos, Alexander, Hutton, \&
  Ward}]{Scott2011}
Scott, A.~E., Stewart, G.~C., Mateos, S., {et~al.} 2011, Monthly Notices of the
  Royal Astronomical Society, 417, 992

\bibitem[{Shemmer {et~al.}(2008)Shemmer, Brandt, Netzer, Maiolino, \&
  Kaspi}]{Shemmer2008}
Shemmer, O., Brandt, W.~N., Netzer, H., Maiolino, R., \& Kaspi, S. 2008, The
  Astrophysical Journal, 682, 81

\bibitem[{Shimizu {et~al.}(2018)Shimizu, Davies, Koss, Ricci, Lamperti, Oh,
  Schawinski, Trakhtenbrot, Burtscher, Genzel, Lin, Lutz, Rosario, Sturm, \&
  Tacconi}]{Shimizu2018}
Shimizu, T.~T., Davies, R.~I., Koss, M., {et~al.} 2018, The Astrophysical
  Journal, 856, 154

\bibitem[{Sobolewska \& Papadakis(2009)}]{Sobolewska2009}
Sobolewska, M.~A. \& Papadakis, I.~E. 2009, Monthly Notices of the Royal
  Astronomical Society, 399, 1597

\bibitem[{Sobral {et~al.}(2013)Sobral, Smail, Best, Geach, Matsuda, Stott,
  Cirasuolo, \& Kurk}]{Sobral2013}
Sobral, D., Smail, I., Best, P.~N., {et~al.} 2013, Monthly Notices of the Royal
  Astronomical Society, 428, 1128

\bibitem[{{Somerville} {et~al.}(2008){Somerville}, {Hopkins}, J., {Robertson},
  \& L.}]{Somerville2008}
{Somerville}, R.~S., {Hopkins}, P.~F., J., C.~T., {Robertson}, B.~E., \& L., H.
  2008, MNRAS, 391, 481

\bibitem[{Stalevski {et~al.}(2012)Stalevski, Fritz, Baes, Nakos, \&
  Popovi{\'{c}}}]{Stalevski2012}
Stalevski, M., Fritz, J., Baes, M., Nakos, T., \& Popovi{\'{c}}, L.~{\v{C}}.
  2012, Monthly Notices of the Royal Astronomical Society, 420, 2756

\bibitem[{Stalevski {et~al.}(2016)Stalevski, Ricci, Ueda, Lira, Fritz, \&
  Baes}]{Stalevski2016}
Stalevski, M., Ricci, C., Ueda, Y., {et~al.} 2016, Monthly Notices of the Royal
  Astronomical Society, 458, 2288

\bibitem[{Suh {et~al.}(2020)Suh, Civano, Trakhtenbrot, Shankar, Hasinger,
  Sanders, \& Allevato}]{Suh2020}
Suh, H., Civano, F., Trakhtenbrot, B., {et~al.} 2020, The Astrophysical
  Journal, 889, 32

\bibitem[{Sun {et~al.}(2015)Sun, Trump, Brandt, Luo, Alexander, Jahnke,
  Rosario, Wang, \& Xue}]{Sun2015}
Sun, M., Trump, J.~R., Brandt, W.~N., {et~al.} 2015, The Astrophysical Journal,
  802, 14

\bibitem[{Taylor(2005)}]{Taylor2005}
Taylor, M.~B. 2005, in Astronomical Society of the Pacific Conference Series,
  Vol. 347, Astronomical Data Analysis Software and Systems XIV, ed.
  P.~{Shopbell}, M.~{Britton}, \& R.~{Ebert}, 29

\bibitem[{Trakhtenbrot {et~al.}(2017)Trakhtenbrot, Ricci, Koss, Schawinski,
  Mushotzky, Ueda, Veilleux, Lamperti, Oh, Treister, Stern, Harrison,
  Balokovic, \& Gehrels}]{Trakhtenbrot2017}
Trakhtenbrot, B., Ricci, C., Koss, M.~J., {et~al.} 2017, MNRAS

\bibitem[{Tranin {et~al.}(2022)Tranin, Godet, Webb, \& Primorac}]{Tranin2022}
Tranin, H., Godet, O., Webb, N., \& Primorac, D. 2022, Astronomy \&
  Astrophysics, 657, A138

\bibitem[{Tristram {et~al.}(2007)Tristram, Meisenheimer, Jaffe, Schartmann,
  Rix, Leinert, Morel, Wittkowski, Röttgering, Perrin, Lopez, Raban, Cotton,
  Graser, Paresce, \& Henning}]{Tristram2007}
Tristram, K. R.~W., Meisenheimer, K., Jaffe, W., {et~al.} 2007, Astronomy \&
  Astrophysics, 474, 837

\bibitem[{{Trump} {et~al.}(2009)}]{Trump2009}
{Trump}, J.~R. {et~al.} 2009, ApJ, 696, 1195

\bibitem[{Urry \& Padovani(1995)}]{Urry1995}
Urry, C.~M. \& Padovani, P. 1995, Publications of the Astronomical Society of
  the Pacific, 107, 803

\bibitem[{Vasudevan \& Fabian(2007)}]{Vasudevan2007}
Vasudevan, R.~V. \& Fabian, A.~C. 2007, Monthly Notices of the Royal
  Astronomical Society, 381, 1235

\bibitem[{Villa-Velez {et~al.}(2021)Villa-Velez, Buat, Theule, Boquien, \&
  Burgarella}]{VillaVelez2021}
Villa-Velez, J.~A., Buat, V., Theule, P., Boquien, M., \& Burgarella, D. 2021,
  Astronomy \& Astrophysics, 654, A153

\bibitem[{Webb {et~al.}(2023)Webb, Carrera, Schwope, Motch, Ballet, Watson,
  Page, Freyberg, Georgantopoulos, Coriat, Barret, Massida, Gupta, Tranin,
  Quintin, Ceballos, Mateos, Corral, Dominguez, Stiele, Traulsen, Pires, Nebot,
  Michel, Pineau, Kuutila, Maggi, Chakroborty, Birchall, Kuin, Akylas, Ruiz,
  Pouliasis, \& Georgakakis}]{Webb2023}
Webb, N.~A., Carrera, F.~J., Schwope, A., {et~al.} 2023, XMM2ATHENA, the H2020
  project to improve XMM-Newton analysis software and prepare for Athena

\bibitem[{Webb {et~al.}(2020)Webb, Coriat, Traulsen, Ballet, Motch, Carrera,
  Koliopanos, Authier, de~la Calle, Ceballos, Colomo, Chuard, Freyberg, Garcia,
  Kolehmainen, Lamer, Lin, Maggi, Michel, Page, Page, Perea-Calderon, Pineau,
  Rodriguez, Rosen, Santos~Lleo, Saxton, Schwope, Tomás, Watson, \&
  Zakardjian}]{Webb2020}
Webb, N.~A., Coriat, M., Traulsen, I., {et~al.} 2020, Astronomy \&
  Astrophysics, 641, A136

\bibitem[{Wu \& Shen(2022)}]{Wu2022}
Wu, Q. \& Shen, Y. 2022, The Astrophysical Journal Supplement Series, 263, 42

\bibitem[{Yang {et~al.}(2022)Yang, Boquien, Brandt, Buat, Burgarella, Ciesla,
  Lehmer, Małek, Mountrichas, Papovich, Pons, Stalevski, Theulé, \&
  Zhu}]{Yang2022}
Yang, G., Boquien, M., Brandt, W.~N., {et~al.} 2022, A\&A [\eprint{2201.03718}]

\bibitem[{Yang {et~al.}(2020)Yang, Boquien, Buat, Burgarella, Ciesla, Duras,
  Stalevski, Brandt, \& Papovich}]{Yang2020}
Yang, G., Boquien, M., Buat, V., {et~al.} 2020, Monthly Notices of the Royal
  Astronomical Society, 491, 740

\bibitem[{Zou {et~al.}(2019)Zou, Yang, Brandt, \& Xue}]{Zou2019}
Zou, F., Yang, G., Brandt, W.~N., \& Xue, Y. 2019, The Astrophysical Journal,
  878, 11

\end{thebibliography}
\bibliographystyle{aa}

\end{document}